\documentclass[sigconf,natbib=true,anonymous=false]{acmart}
\usepackage{bm}
\usepackage{multirow}
\usepackage[normalem]{ulem}
\useunder{\uline}{\ul}{}
\usepackage{algorithmic}
\usepackage[ruled, linesnumbered]{algorithm2e}
\usepackage{booktabs} % For formal tables
\usepackage{amsmath}

\usepackage{amssymb}
\usepackage{xcolor}
\usepackage{multirow}
\usepackage{graphicx}
\usepackage{amsfonts}
\usepackage{balance}
\usepackage{enumitem}
\usepackage{float}                  
\usepackage{subfig}                 
\usepackage{overpic}  
\usepackage{graphicx}

\begin{document}

\title{Double Correction Framework for Denoising Recommendation}

\author{Zhuangzhuang He}
\authornote{This work is produced during a research intern at Tsinghua University.}
\authornote{Equal Contribution.}
\affiliation{%
  \institution{
  School of Computer Science and Information Engineering,
  Hefei University of Technology}
  \country{}
}
\email{hyicheng223@gmail.com}

\author{Yifan Wang}
\authornotemark[2]
% \authornote{}
\affiliation{%
  \institution{
  Department of Computer Science and Technology, Tsinghua University}
  \country{}
}
\email{yf-wang21@mails.tsinghua.edu.cn}

\author{Yonghui Yang}
\affiliation{%
  \institution{School of Computer Science and Information Engineering, Hefei University of Technology}
  \country{}
}
\email{yyh.hfut@gmail.com}

\author{Peijie Sun}
\affiliation{%
  \institution{Department of Computer Science and Technology, Tsinghua University}
  \country{}
}
\email{sun.hfut@gmail.com}

\author{Le Wu}
\authornote{Corresponding authors.}
\affiliation{%
  \institution{School of Computer Science and Information Engineering, Hefei University of Technology}
\institution{Institute of Dataspace,\\ Hefei Comprehensive National Science Center}
  \country{}
}
\email{lewu.ustc@gmail.com}

\author{Haoyue Bai}
\affiliation{%
  \institution{
  School of Computer Science and Information Engineering,
  Hefei University of Technology}
  \country{}
}
\email{baihaoyue621@gmail.com}

\author{Jinqi Gong}
\affiliation{%
  \institution{Department of Mathematics, University of Macau}
  \country{}
}
\email{eggmangong@gmail.com}

\author{Richang Hong}
\affiliation{%
  \institution{School of Computer Science and Information Engineering, Hefei University of Technology}
  \country{}
}
\email{hongrc.hfut@gmail.com}

\author{Min Zhang}
\authornotemark[3]
\affiliation{%
  \institution{Department of Computer Science and Technology, Tsinghua University}
  \country{}
}
\email{z-m@tsinghua.edu.cn}

\renewcommand{\shortauthors}{Zhuangzhuang et al.}

\begin{abstract}
As its availability and generality in online services, implicit feedback is more commonly used in recommender systems. However, implicit feedback usually presents noisy samples in real-world recommendation scenarios~(such as misclicks or non-preferential behaviors), which will affect precise user preference learning. To overcome the noisy sample problem, a popular solution is based on dropping noisy samples in the model training phase, which follows the observation that noisy samples have higher training losses than clean samples. Despite the effectiveness, we argue that this solution still has limits. (1) High training losses can result from model optimization instability or hard samples, not just noisy samples. (2) Completely dropping of noisy samples will aggravate the data sparsity, which lacks full data exploitation.

To tackle the above limitations, we propose a \textit{\textbf{D}ouble \textbf{C}orrection \textbf{F}ramework for Denoising Recommendation} (\textbf{DCF}), which contains two correction components from views of more precise sample dropping and avoiding more sparse data. In the sample dropping correction component, we use the value of the sample loss over time to determine whether it is noise or not, increasing stability. Instead of averaging directly, we use the damping function to reduce the bias effect of outliers. Furthermore, due to the higher variance exhibited by hard samples, we derive a lower bound for the loss through concentration inequalities to identify and reuse hard samples. In progressive label correction, we iteratively re-label highly deterministic noisy samples and retrain them to further improve performance. Finally, extensive experimental results on three datasets and four backbones demonstrate the effectiveness and generalization of our proposed framework. Our code is available at \href{https://github.com/bruno686/DCF}{https://github.com/bruno686/DCF}.
\end{abstract}



% \keywords{recommendation, denoising, implicit feedback}

\maketitle

\section{Introduction}
\label{introduction}
Recommender systems often use implicit feedback (e.g., click, watch, and purchase) to learn user interests and recommend items, where users' implicit feedback is thought to reflect users' true preferences~\cite{sun1,sun2,sun3, hui1,bai1,bai2,cai1,wang1,wang2}. However, here are some exceptions. For example, a user may purchase and return an item or click on a video but the dwell time is very short, thus the purchasing and clicking behavior may not represent users' interests, which is referred to \textit{noisy interaction} by recent research~\cite{DeCA, T-CE}. 
Recent research~\cite{DeCA, T-CE} refer to this type of interaction as~\textit{noisy interaction}.
Treating these noisy interactions as clean interactions may get suboptimal performance~\cite{T-CE, DeCA, post-click}. Therefore, how to eliminate the adverse effects of noisy interactions has attracted a broad interest from the research community.

\begin{figure}[t]
	\centering
	\subfloat[Illustration of unstable losses (dotted line represents the mean of losses).]{\includegraphics[width=.51\linewidth]{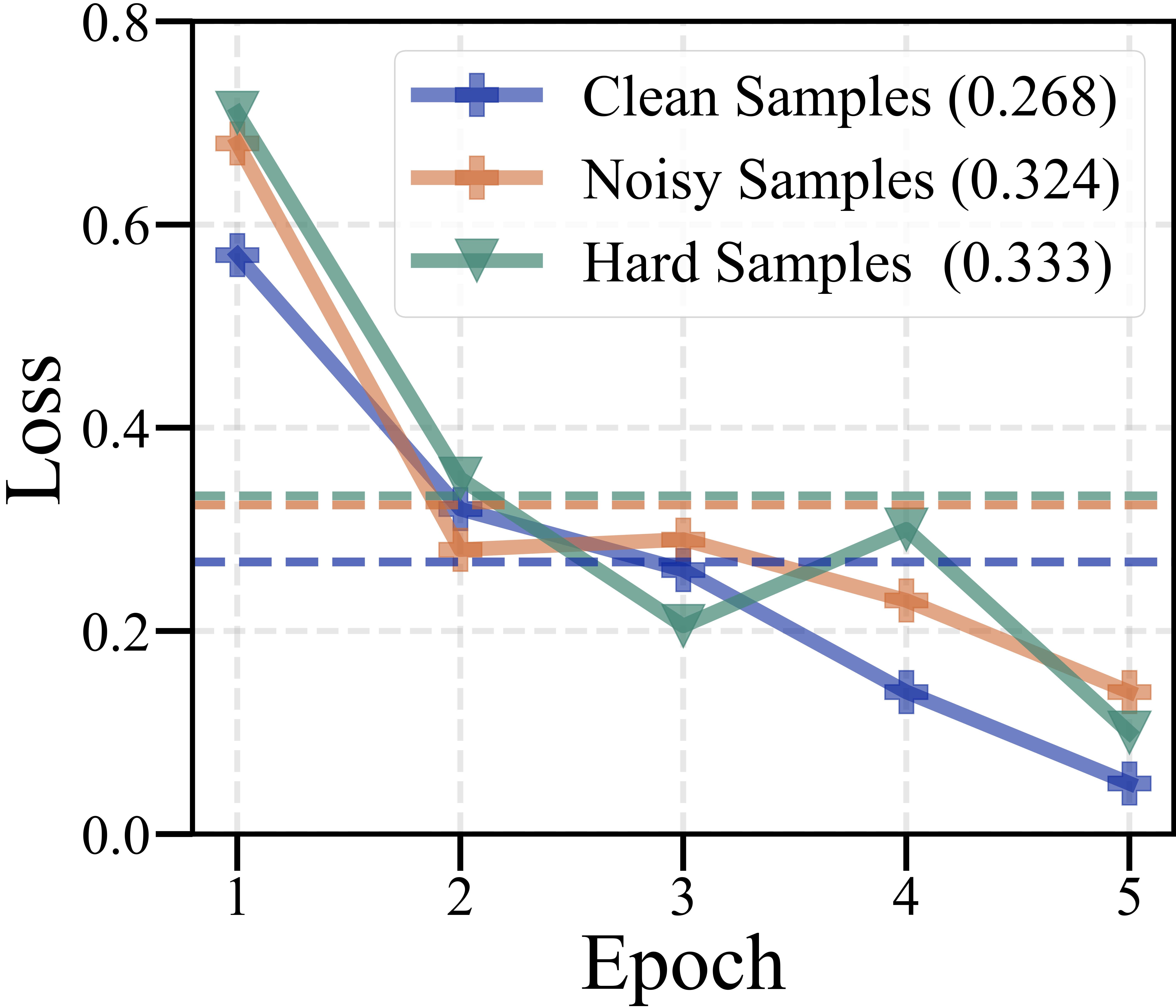}}\hspace{5pt}
	\subfloat[Different strategies in their ideal cases.]{\includegraphics[width=.46\linewidth]{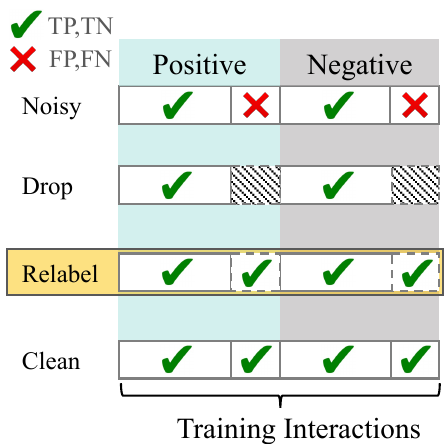}}
   \caption{(a)~\uline{Illustration of unstable losses}. We observe that clean samples do not exhibit low loss in every epoch. Similarly, noisy samples do not always exhibit high loss. Also, hard samples exhibit high losses. However, the noisy samples can be identified from the perspective of mean loss. (b)~\uline{Different strategies in their ideal cases}. Explain the difference between our relabeling strategy and other strategies under ideal conditions. Here, the TP, TN of T stands for true. Similarly, F stands for false.}
\vspace{-2mm}
\label{intro}
\end{figure}

Although some noisy interactions may be detected by multiple behaviors (e.g., explicit feedback), we cannot guarantee that other behaviors validate every interaction due to the sparsity of user behavior. Additionally, the request for explicit information is expensive as it may hurt the user experience. Thus, it is important to discover noisy interactions based on their unique data pattern,  which has also received much attention in recent years. A commonly observed pattern, which is also the basis of most current work~\cite{T-CE, AutoDenoise-reinforce, SGDL}, is that the loss values for noisy interactions are generally higher than for clean ones during training. Hence, a straightforward solution is to use loss values as a distinguishing feature to process samples with high losses. For instance,~\cite {T-CE, AutoDenoise-reinforce} directly and dynamically drops samples with high losses during training, achieving notable results. \looseness=-1

Despite the notable performance, we argue that current methods~\cite{T-CE, AutoDenoise-reinforce, SGDL} have two limitations that may hinder effectiveness.
(1) \uline{They ignore that losses may not be highly correlated with noise}, i.e., the instability of optimization and the hardness of clean interactions. 
Firstly, the instability in the optimization process may lead to sharp changes in the immediate loss values of the sample, which may show a contrary phenomenon to the above basis. As illustrated in Fig.~\ref{intro}~(a), noisy samples also show lower training loss sometimes, and vice versa, which leads to incorrect dropping and may hurt the performance. Secondly, hard samples also typically show high loss and could be dropped along with noisy samples.
Abandoning these hard samples may lead to suboptimal performance as they are informative and beneficial for recommendation performance ~\cite{hard_advantage}. 
(2) \uline{Simply dropping samples may lead to more sparse data.} As illustrated in Fig.~\ref{intro}~(b), although the noisy interactions are mislabeled, they are still part of the training space. Merely dropping these noisy samples leads to sample wastage, as well as potentially causing the training space to be inconsistent with the ideal clean space.

To address these limitations, we explore potential solutions and the corresponding challenges. For the first point, we attribute the low correlation between loss and noise to limited observations of current model predictions and the neglect to consider hard samples. Hence, we expect to stabilize the model's predictions by extending the observation interval and aggregating loss values from multiple training iterations. 
In addition, high-loss samples might be hard samples beneficial for training. Therefore, we aim to identify and retain these hard samples to enhance recommendation performance. However, how to efficiently and simply identify hard samples is still challenging. For the second point, we argue that even if noisy samples, have corresponding correct labels and cannot be merely dropped from the sample space, which would result in more sparse space. Hence, we would like to relabel and reintroduce part of noisy samples that are highly determined to be noise into the training process. And we illustrate in Fig.~\ref{intro}(b) how relabeling addresses the issue of sample waste caused by drops. However, there is a practical challenge in determining which samples need to be relabelled and the proportion of relabelling during model training.

In this paper, we propose a \uline{\textbf{D}}ouble \uline{\textbf{C}}orrection \uline{\textbf{F}}ramework for Denoising Recommendation (\textbf{DCF}). The core of this framework is sample dropping correction and progressive label correction, which are designed for the above two limitations respectively.
Sample dropping correction combines confirmed loss calculation and cautious hard sample search to accurately drop noisy samples and retain hard samples. Specifically, we calculate the mean loss of samples over a time interval and robustly compute each loss value using damping functions to mitigate the effects of occasional outliers. Inspired by~\cite{variance}, hard samples have a higher loss variance than clean and noisy samples. In other words, the loss value of hard samples has a lower bound in terms of the whole training process. Therefore, we use concentration inequalities to derive confidence intervals for each sample's loss value. Then we use the lower bound of the calculated confidence interval as a hard sample search criterion.
In this way, hard samples are retained for training instead of being dropped.
In the progressive label correction component, we believe the stability of the model optimization process increases gradually~\cite{stable}, so the model's relabeling strategy should adapt to this characteristic. Specifically, we initially relabel a small fraction of the samples and progressively increase the proportion of relabeling as training proceeds. Note that we still drop samples with high loss values and just relabel a fraction of the samples with highly determined to be noisy.

To summarize, our main contributions are as follows:
% \vspace{-10pt}
\begin{itemize}[leftmargin=*]
  \item We analyze two limitations of the loss-based dropping recommendations, (1)~loss values are not highly correlated with noise, and (2)~complete dropping of noisy samples, which leads to more sparse data space.  
  \item We design two correction components to enhance the effectiveness of denoising recommendations from more precise sample dropping and reusing noisy sample perspectives, respectively.
  \item We experiment with our proposed DCF of four backbones on three popular datasets. And extensive experimental results demonstrate the effectiveness and generalization of our framework.
\end{itemize}
%%+

\begin{figure*}[th]
\centering
\includegraphics[width=0.8\linewidth]{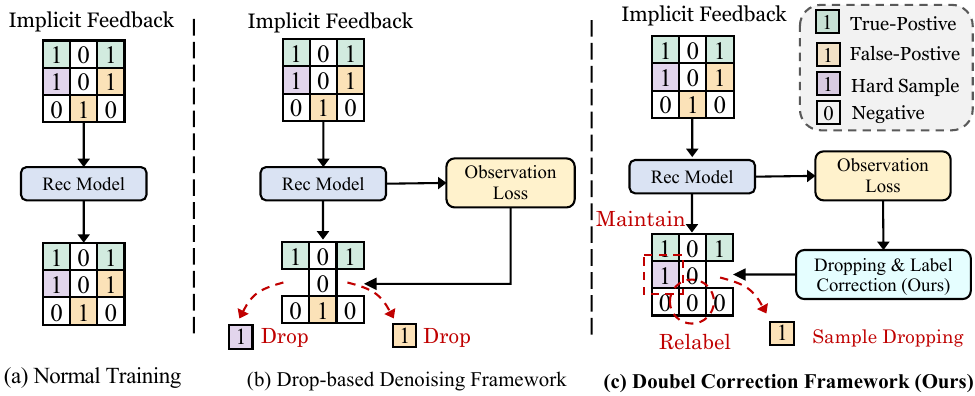}
\caption{Illustration comparison: (a) Normal training model without denoising, (b) Denoising model with drop strategy, (c) Double correction framework for denoising recommendation~(Ours).}
\label{zongti}
\vspace{-3mm}
\end{figure*}

\section{task description}
We first give a formal description of the denoising recommendation task.
We use $u \in \mathcal{U}$ to denote users, $p \in \mathcal{P}$ to denote the item, and observed interaction matrix $\tilde{\mathbf{Y}} \in \mathbb{R}^{|\mathcal{U}| \times|\mathcal{P}|}$. 
And where $\tilde{y}_{ui} = 1$ means that the user interacted with the item, and $\tilde{y}_{ui} = 0$ means no interaction. In previous work, the default assumption is that whenever $\tilde{y}_{ui} = 1$, it means that the user likes the item. However, user interactions may be due to various noises~(i.e., clicking on a video by mistake or clicking on a video for a very short time to exit), resulting in users not liking the interacted item.
Therefore, the target of denoising recommendations is to develop a model~$f$ with parameters $\theta_f$ to learn noise-free representation of users and items from noisy data~$\tilde{\mathbf{Y}}$.
The training of denoising recommendation model is formulated as follows:
$$
   \theta_f^*=\underset{\theta_f}{\arg \min } \mathcal{L}_{\text {rec}}(\tilde{\mathbf{Y}}), 
$$
where $\mathcal{L}_{\text {rec}}$ is the recommendation loss, such as BPR loss~\cite{diffnet} and BCE loss~\cite{T-CE}, and $\theta_f^*$ is the optimal parameters of $f$. 
Here, following previous denoising recommendation works~\cite{T-CE,AutoDenoise-reinforce,DeCA}, we use the BCE loss as an instantiation of $\mathcal{L}_{rec}$ :
$$  
\mathcal{L}_{rec}=-\underset{(u, p) \sim P_{\tilde{\mathbf{Y}}}} {\mathbb{E}}\left[\tilde{y}_{u,p} \cdot \log \left(\hat{y}_{u,p}\right)+\left(1-\tilde{y}_{u,p}\right) \cdot \log \left(1-\hat{y}_{u,p}\right)\right],
$$
where $P_{\tilde{\mathbf{Y}}}(\cdot)$ denotes the distribution established over the interaction data. The $\hat{y}_{u,p}$ represents the prediction made by the model for the user's preference. In addition, the loss of one interaction is $\ell$.

\section{THE PROPOSED FRAMEWORK}
\subsection{Overview}
\begin{figure}[t]
\centering
\includegraphics[width=0.9\linewidth]{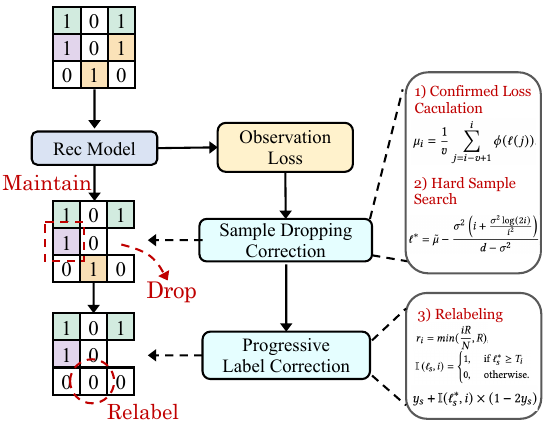}
\caption{Details of Loss \& Label Correction.}
\label{details}
\vspace{-3mm}
\end{figure}

Our framework DCF~(as shown in Fig.~\ref{zongti}~(c)) is developed on a common observation: noisy samples usually incur high losses \cite{T-CE}.
However, on the one hand, the model optimization is usually unstable, and hard samples also show high losses; on the other hand, some methods~\cite{T-CE, AutoDenoise-weight} directly drop samples with high loss values for simplicity. Such actions may lose feature information and degrade performance.
To mitigate these limitations, we design two components~(as shown in Fig.~\ref{details}): sample dropping correction and progressive label correction.
The first component, sample dropping correction, aims to correct unstable loss values as model random initialization parameters and unstable optimization processes. In addition, we also calculate the confidence intervals for each sample loss and use the lower bound as our dropping strategy. This method aims to retain hard samples because hard samples enhance model performance by exposing edge cases and revealing more profound insights into data patterns~\cite{hard_advantage}.
Additionally, for those samples that are highly determined to be noise, the second component, progressive label correction, relabels these and adds them to the next training iteration to improve performance further.
Based on these two components, our framework mitigates the adverse effects of noisy interactions more effectively. The complete algorithmic process is at Algorithm~\ref{alg:bemap}.

% \subsection{Hard Samples Search}
\subsection{Sample Dropping Correction}
Due to the instability of the optimization process and randomness of the initial weights. 
As shown in Fig.~\ref{intro}~(a), true positive interactions may exhibit high loss at some time points, while noisy interactions may exhibit low ones. Fortunately, we also observe that the sample loss can be more stable by replacing the loss value at a single time point with the mean loss over time. Thus, inspired by this, we estimate the sample loss using a confirmed approach. Besides, samples that perform with high loss may also be hard samples, which can enhance the model performance~\cite{hard_advantage}. In a word, our goal in this subsection is to drop the noisy samples more accurately, find the hard samples, and retain them.

\subsubsection{\textbf{Confirmed Loss Calculation}}
Relying only on empirical losses calculated from a single training iteration may incorrectly distinguish between noise and clean, thus degrading performance. Therefore, we consider the loss values across previous training iterations to mitigate misjudge.
Specific practice is to use the mean of losses at different iterations.
\begin{equation}
\mu_i=\frac{1}{v} \sum_{j=i-v+1}^i \ell (j),
\end{equation}
where we calculate the mean loss value $\mu_i$ over $v$ time intervals prior to the $i$-th training iteration (including the $i$-th). And if the current iteration is less than $v-1$ times, $\mu_i$ is the mean of all losses up to the current iteration $i$.
However, during the training process, the model may encounter extreme loss values. Although this possibility is unlikely, to avoid negative impacts when calculating means, we need to take defensive measures for robust computation.
Thus, instead of calculating the mean directly, we robustly process each loss. After that, we obtain the mean as follows:
\begin{equation}
\mu_i=\frac{1}{v} \sum_{j=i-v+1}^i \phi(\ell (j)),
\label{mu}
\end{equation}
where, the non-decreasing damping function $\phi(\cdot)$ is used to calculate mean values more robustly. The specific forms are as follows:
$
\phi(\ell )=\log \left(1+\ell +\ell ^2 / 2\right).
$

The advantages of the damping function include: (1) \uline{Reduced impact of extremes.} 
The damping function grows slowly at very large loss values because it is logarithmic. These large values may be outliers due to unstable model predictions from noisy samples. Therefore, the effect of extreme values is reduced when the mean is calculated afterward. (2) \uline{Approximate linearity at small values.} At small values, $\phi(\cdot)$ can be approximated as linear function. This means that near small values, $\phi(\cdot)$ does not have side effects on the loss values, preserving the original.

\subsubsection{\textbf{Cautious Hard Samples Search}}
We argue that dropping samples with higher mean loss still leaves room for potential improvements. This is because a higher mean sample loss does not necessarily indicate noisy samples; they could be hard samples.~\cite{T-CE} point out that the hard samples contain more information than the simple ones. Thus we aim to search and retain hard samples. Our search strategy is based on the observation~\cite{variance} that hard samples exhibit higher variance in loss values during training.
Consequently, we compute the lower confidence interval bound for the loss by \textit{Concentration Inequalities}~\cite{fang}. 
Because, for noisy samples, the lower bound of the loss confidence interval usually closely matches the actual loss value. However, for hard samples with high variance, the confidence lower bound tends to be lower. Therefore, in our sample dropping and subsequent relabeling strategy, we use the lower bound as our criterion rather than the loss value itself. We define the form of centralized inequality specified below.

\vspace{5pt}
\noindent
\textbf{Theorem 1.} \textit{Let $Z_n=\left\{z_1, \cdots, z_n\right\}$ be an observation set with mean $\mu_z$ and variance $\sigma^2$. By exploiting the non-decreasing damping function $\phi(z)=\log \left(1+z+z^2 / 2\right)$. For any $\epsilon>0$, we have}
\begin{equation}
\left|\frac{1}{n} \sum_{i=1}^n \phi\left(z_i\right)-\mu_z\right| \leq \frac{\sigma^2\left(n+\frac{\sigma^2 \log \left(\epsilon^{-1}\right)}{n^2}\right)}{n-\sigma^2},
\end{equation}
with probability at least $1-2 \epsilon$.

There are other methods to filter out the hard samples, such as gradient~\cite{gradient} and cluster~\cite{cluster}. Compared to these methods, our framework has the following advantages. (1)~\uline{Intuition}. Lower bound, as a basic statistical concept, can intuitively reflect the degree of data dispersion. For hard samples, their loss values fluctuate more during the training process, and thus the lower bound is lower. (2)~\uline{Computational efficiency}. The low complexity of computing lower bound is suitable for large-scale datasets. Compared to other selection methods, the lower bound of sample loss provides a simple and efficient strategy. \looseness=-1

Let, $\epsilon=\frac{1}{2i}$, the following loss lower bound $\ell^{\ast}$ form is obtained after derivation. The detailed derivation process is in Appendix~\ref{PROOF}.
\begin{equation}
\ell^{\ast}=\tilde{\mu}-\frac{\sigma^2\left(i+\frac{\sigma^2 \log (2 i)}{i^2}\right)}{d-\sigma^2},
\label{low bound1}
\end{equation}
where $d$ represents the number of times a sample has not been dropped. Intuitively, the smaller $d$ is, the lower the confidence interval bound will be. We prioritize reincorporating samples with lower confidence interval bounds (i.e., higher variance) into training. In addition, the search intensity is retained through an adjustment factor, $\sigma^2$, that reduces the noisy samples' adverse effects and increases the model performance. It is important to note that we are not selecting samples based on their loss values but rather based on the lower bound.

Through the sample dropping correction strategy, we achieve higher confirmed sample loss values and effectively search and retain hard samples, further enhancing the model performance.

\begin{algorithm}[t]
	\SetKwInOut{Input}{Input}
	\SetKwInOut{Output}{Output}
	\Input{Training set $D$, Maximum Epochs $N$, Time interval $v$, Relabel Ratio $R$;}
	\Output{Trained Model M$\mathcal{M}$.}
	Initialize model $\mathcal{M}$ with random weights\; 
	Initialize empty loss history $L$ for each sample in $D$\; 
	\For{$i = 1$ \KwTo N}{
	    \For{each mini-batch B from D}{
                \tcp{Sample Dropping Correction}
	        \For{each sample s in B}{
                    Compute loss $\ell$ using model $\mathcal{M}$\;
                    Append $\ell$ to loss history $L[s]$\;
                    \eIf{$length(L[s]) > v$}{Remove samples in $L[s]$ with time interval greater than $v$\;}{
                    Compute mean loss $\mu$ using Eq.~\eqref{mu}\;
                    Compute lower bound $\ell$* using Eq.~\eqref{low bound1}\;
                    }
                }
                Update model $\mathcal{M}$ using lower bound $\ell$*\;
	    }
            \tcp{Progressive Label Correction}
            Compute relabel ratio $r$ using Eq.~\eqref{relabel ratio}\;
            \For{each sample $s$ in D}{
                \eIf{$\ell$* >= $T_i$ ($T_i$ is computed based on r and $\ell$*)}{Flip label $y_s$ using Eq.~\eqref{new label}\;}{Keep label $y_s$\;}
            }
            Training the model with the corrected samples at the next epoch;
	}
	\Return Trained Model $\mathcal{M}$;
	\caption{Our proposed $\mathsf{DCF}$.}
	\label{alg:bemap}
\end{algorithm}

\subsection{Progressive Label Correction}
Compared to directly dropping noisy samples. We argue that the small fraction of samples with the highest loss~\footnote{For simplicity in description, we still use the loss to represent the lower bound of the loss value. But we have the notational distinction between $\ell$ and $\ell$*} can be leveraged in a relabeling way. Relabeling noisy labels preserves samples rather than dropping them, providing more training samples. This is valuable for sparse recommendation datasets, which makes the entire training sample space and test set space more consistent. Moreover, relabeling is a simple operation that does not require complex reweighting strategies or other processing techniques.
Thereby further training the model adequately to improve its performance. In addition, since the predictive stability of the model improves incrementally during training, we argue it is reasonable to relabel more in the later stages and less in the earlier stages. From the beginning to the end of the training, we progressively increased the percentage of relabeling.

We first define the percentage of progressive relabeling. Our relabel ratio $r_i$ grows as epoch $i$ increases. It is defined as follows:
\begin{equation}
r_i = min(\frac{i R}{O}, R),
\label{relabel ratio}
\end{equation}
where $O$ denotes that the flipping rate remains constant after the $O$-th epoch, and $R$ is the percentage we end up relabeling. This way, we get each epoch's relabeled ratio $r_i$. And the corresponding loss threshold of the samples to be labeled is $T_i=l[\lfloor B(1-r_i) \rfloor]$. $\lfloor \cdot \rfloor$ is a downward rounding operation.
If the $\ell_i$ is greater than $T_i$, the indicator $\mathbb I(\ell_s, i)$ is $1$.
\begin{equation}
\mathbb I\left(\ell_s^{\ast}, i\right)= \begin{cases}1, & \text { if } \ell_s^{\ast} \geq T_i \\ 0, & \text { otherwise. }\end{cases}
\end{equation}
Up to this point, we know which samples we should flip the labels on. We flip labels in the dataset so that the flipped
labels are also used at subsequent training epochs.
\begin{equation}
y_s^{\prime}=y_s+\mathbb I(\ell_s^{\ast}, i) \times(1-2 y_s),
\label{new label}
\end{equation}
where we relabel the original label from $y_s$ to $y_s^{\prime}$. After careful processing, we relabel these noisy labels to ensure they are closer to the true ones. This step not only enhances the data quality but also provides more accurate labels for subsequent model optimization, which is expected to improve the effectiveness. 

Through the progressive label correction strategy, we have avoided sample waste, thereby providing more data for the precise modeling of users and items.

\subsection{Model Discussion}
This section compare different recommendation models based on their space and time complexities. The comparison is summarized in Table~\ref{model complexity}. In addition, we discuss the improved features of $\mathsf{DCF}$ to the previous dropping approach T-CE~\cite{T-CE}.

\vspace{5pt}
\noindent
\textbf{Space Complexity.} 
The space complexity of each model is determined by the number of parameters, denoted as $M$. The base model and T-CE model have the same space complexity of $M$ since T-CE does not add any new parameters or structures. However, BOD includes a weight generator with an encoder layer $EN\in\mathbb{R}^{2d \times d_{g}}$ and a decoder layer $DE\in\mathbb{R}^{d_{g} \times 1}$, where $d$ is the embedding size of user and item in the recommendation model, and $d_{g}$ is the hidden layer size of the generator. On the other hand, $\mathsf{DCF}$ introduces additional computations like confidence bounds of the loss but does not add new model parameters, meaning its space complexity is also $M$.

\vspace{5pt}
\noindent
\textbf{Time Complexity.} 
Regarding time complexity, the base model computes the loss directly in $\mathcal O(n)$ time, where $n$ is the number of samples. T-CE sorts the computed loss, resulting in a time complexity of $\mathcal O(n \log n)$. BOD involves bi-level optimization, leading to a time complexity of $\mathcal O\left(n+d\left(d_g\right)+\left(d_g\right)\right)$. Our proposed method includes multiple steps like computing confidence bounds, soft processing, and sorting. Sorting is the dominant factor here, leading to a time complexity of $\mathcal O(n \log n)$.

\vspace{-7pt}
\begin{table}[H]
  \caption{Model complexity comparison.}\vspace{-1em}
  \small
  \label{model analysis}
\begin{tabular}{ccc}
\toprule
\textbf{Model}            & \textbf{Space Complexity} & \textbf{Time Complexity} \\
\midrule
$\mathsf{Base}$~\cite{deepl} & $M$ & $\mathcal O(n)$ \\ 
$\mathsf{T}$-$\mathsf{CE}$~\cite{T-CE} & $M$    & $\mathcal O(n \log n)$ \\ 
$\mathsf{BOD}$~\cite{BOD} & $M+EN+DE$  & $\mathcal O\left(n+d\left(d_g\right)+\left(d_g\right)\right)$ \\ 
$\mathsf{\textbf{DCF (Ours)}}$ & $M$     & $\mathcal O(n \log n)$ \\ 
\bottomrule
\end{tabular}
\label{model complexity}
\end{table}

\vspace{-7pt}
\noindent
\textbf{In-depth Comparison.} The T-CE method overlooks the potential of hard samples, possibly hindering model performance. Our cautious hard sample search, however, recognizes the value in these samples. By focusing on the lower bound of loss rather than just mean loss, we can better differentiate between noise and hard challenges. This approach ensures we capture the full learning spectrum, addressing T-CE's limitation.

\section{Experiments}
To demonstrate the effectiveness and generalization of our proposed method, we compare the results of our DCF with state-of-the-art recommendation models on four backbones and three real-world datasets.
We aim to answer the following research questions:
\begin{itemize}[leftmargin=*]
  \item \textbf{RQ1 }: How does our proposed method perform compared to normal training and other state-of-the-art denoising methods?  
  \item \textbf{RQ2 }: How does each component in the DCF contribute to the overall performance and the impact of hyperparameters? 
  \item \textbf{RQ3 }: Do the hard samples we search for improve other model performance compared to random sampling? 
  \item \textbf{RQ4 }: Does our progressive strategy provide better results compared to the fixed flip ratio strategy?
\end{itemize}

\vspace{-7pt}
\begin{table}[H]
\begin{center}
\caption{Statistics of datasets.}
\vspace{-7pt}
\label{tab:dataset statistics}
\begin{tabular}{crrrc}
  \toprule
  \textbf{Dataset} & \textbf{\#Users} & \textbf{\#Items} & \textbf{\#Interactions} & \textbf{Sparsity} \\
  \midrule
  Adressa & 212,231 & 6,596 & 419,491 & 0.99970 \\
%   \hline
  MovieLens & 943 & 1,682 & 100,000 & 0.93695 \\
%   \hline
  Yelp & 45,548 & 57,396 & 1,672,520 & 0.99936 \\
  \bottomrule
\end{tabular}
\end{center}
\vspace{-0.5cm}
\end{table}

\begin{table*}[h]
\LARGE
\renewcommand\arraystretch{1}
\caption{Performance comparison of different denoising methods on the robust recommendation. The highest scores are in bold, and the \uline{runner-ups} are with underlines. R and N refer to Recall and NDCG, respectively. A significant improvement over the runner-up is marked with * (i.e., two-sided t-test with $p \textless 0.05$) and ** (i.e., two-sided t-test with $0.05 \leq p \textless 0.1$).}\vspace{-1em}
\centering

\label{overrall compa}
\vspace{0.25cm}
\resizebox{0.95\textwidth}{!}{
\begin{tabular}{@{}cc|llll|llll|llll@{}}
\toprule[1.2pt]

\multicolumn{2}{c}{\textbf{Dataset}}& \multicolumn{4}{|c}{\textbf{Adressa}}& \multicolumn{4}{|c}{\textbf{MovieLens}}& \multicolumn{4}{|c}{\textbf{Yelp}}   \\ 
\midrule

\multicolumn{1}{c}{\textbf{Base Model}}& \textbf{Method} & \multicolumn{1}{c}{\textbf{R@5}}& \multicolumn{1}{c}{\textbf{R@20}}& \multicolumn{1}{c}{\textbf{N@5}}  & \multicolumn{1}{c}{\textbf{N@20}} & \multicolumn{1}{|c}{\textbf{R@5}}& \multicolumn{1}{c}{\textbf{R@20}} & \multicolumn{1}{c}{\textbf{N@5}} & \multicolumn{1}{c}{\textbf{N@20}} & \multicolumn{1}{|c}{\textbf{R@5}}& \multicolumn{1}{c}{\textbf{R@20}} & \multicolumn{1}{c}{\textbf{N@5}}& \multicolumn{1}{c}{\textbf{N@20}}
 \\ 
\midrule
\midrule
\multicolumn{1}{c}{\multirow{7}{*}{GMF}}      & Normal & \multicolumn{1}{l}{0.1257} & \multicolumn{1}{l}{0.2126} & \multicolumn{1}{l}{0.0908} & 0.1210 & \multicolumn{1}{l}{0.0355} & \multicolumn{1}{l}{0.1196} & \multicolumn{1}{l}{0.0482} & 0.0715  & \multicolumn{1}{l}{0.0149} & \multicolumn{1}{l}{0.0431} & \multicolumn{1}{l}{0.0152} & 0.0243 \\
\multicolumn{1}{c}{} & WBPR   & \multicolumn{1}{l}{0.1258} & \multicolumn{1}{l}{0.2131} & \multicolumn{1}{l}{0.0912} & 0.1212 & \multicolumn{1}{l}{0.0357} & \multicolumn{1}{l}{0.1199} & \multicolumn{1}{l}{0.0486} & 0.0718 & \multicolumn{1}{l}{0.0148} & \multicolumn{1}{l}{0.0434} & \multicolumn{1}{l}{0.0155} & 0.0246\\
\multicolumn{1}{c}{} & WRMF & \multicolumn{1}{l}{\uline{0.1263}} & \multicolumn{1}{l}{0.2132} & \multicolumn{1}{l}{0.0911} & 0.1214 & \multicolumn{1}{l}{0.0363} & \multicolumn{1}{l}{\uline{0.1201}} & \multicolumn{1}{l}{0.0491} & 0.0722 & \multicolumn{1}{l}{0.0144}    & \multicolumn{1}{l}{0.0446}  & \multicolumn{1}{l}{0.0140} &  0.0242\\
\multicolumn{1}{c}{} & T-CE   & \multicolumn{1}{l}{0.1251} & \multicolumn{1}{l}{\uline{0.2155}} & \multicolumn{1}{l}{\uline{0.0913}} & \uline{0.1228} & \multicolumn{1}{l}{\uline{0.0374}} & \multicolumn{1}{l}{\textbf{0.1202}} & \multicolumn{1}{l}{\uline{0.0509}} & \uline{0.0738} & \multicolumn{1}{l}{0.0143} & \multicolumn{1}{l}{\uline{0.0447}} & \multicolumn{1}{l}{0.0143} & 0.0244\\
\multicolumn{1}{c}{} & DeCA   & \multicolumn{1}{l}{0.1232}    & \multicolumn{1}{l}{0.2147}    & \multicolumn{1}{l}{0.0866}    &  0.1181  & \multicolumn{1}{l}{0.0364}    & \multicolumn{1}{l}{0.1075}    & \multicolumn{1}{l}{0.0423} & 0.0631  & \multicolumn{1}{l}{0.0141}    & \multicolumn{1}{l}{0.0442}    & \multicolumn{1}{l}{0.0139}    &   0.0236  \\ 
\multicolumn{1}{c}{} & BOD   & \multicolumn{1}{l}{0.1237}    & \multicolumn{1}{l}{0.2153}    & \multicolumn{1}{l}{0.0892}    &  0.1193    & \multicolumn{1}{l}{0.0366}    & \multicolumn{1}{l}{0.1083}    & \multicolumn{1}{l}{0.0474} & 0.0685 &   \multicolumn{1}{l}{\uline{0.0151}} & \multicolumn{1}{l}{0.0436} & \multicolumn{1}{l}{\uline{0.0157}} & \uline{0.0247} \\
\cmidrule(l){2-14} 
\multicolumn{1}{c}{} & $\mathsf{\textbf{DCF (Ours)}}$    & \multicolumn{1}{l}{\textbf{0.1296}*} & \multicolumn{1}{l}{\textbf{0.2183}*} & \multicolumn{1}{l}{\textbf{0.0938}*} &  \textbf{0.1254}*  & \multicolumn{1}{l}{\textbf{0.0427}*} & \multicolumn{1}{l}{0.1175} & \multicolumn{1}{l}{\textbf{0.0543}*} &  \textbf{0.0743}*  & \multicolumn{1}{l}{\textbf{0.0155}**} & \multicolumn{1}{l}{\textbf{0.0458}*} & \multicolumn{1}{l}{\textbf{0.0158}} & \textbf{0.0257}* \\ 

\midrule
\midrule
\multicolumn{1}{c}{\multirow{7}{*}{NeuMF}}   & Normal & \multicolumn{1}{l}{0.1909} & \multicolumn{1}{l}{0.3078} & \multicolumn{1}{l}{0.1427} & 0.1851 & \multicolumn{1}{l}{
\uline{0.0439}} & \multicolumn{1}{l}{0.1084} & \multicolumn{1}{l}{\uline{0.0516}} & 0.0724 & \multicolumn{1}{l}{0.0123} & \multicolumn{1}{l}{0.0386} & \multicolumn{1}{l}{\uline{0.0123}} & 0.0210 \\
\multicolumn{1}{c}{} & WBPR   & \multicolumn{1}{l}{0.1903} & \multicolumn{1}{l}{0.3082} & \multicolumn{1}{l}{\uline{0.1428}} & 0.1848 & \multicolumn{1}{l}{0.0426} & \multicolumn{1}{l}{0.1132} & \multicolumn{1}{l}{0.0504} & \uline{0.0735} & \multicolumn{1}{l}{0.0104}    & \multicolumn{1}{l}{0.0384}    & \multicolumn{1}{l}{0.0108}  &  0.0193\\
\multicolumn{1}{c}{} & WRMF   & \multicolumn{1}{l}{\uline{0.1922}} & \multicolumn{1}{l}{\uline{0.3084}} & \multicolumn{1}{l}{0.1424} & \uline{0.1852} & \multicolumn{1}{l}{0.0418} & \multicolumn{1}{l}{\uline{0.1180}} & \multicolumn{1}{l}{0.0512} & 0.0729 & \multicolumn{1}{l}{0.0119} & \multicolumn{1}{l}{0.0378} & \multicolumn{1}{l}{0.0121} & 0.0198\\
\multicolumn{1}{c}{} & T-CE   & \multicolumn{1}{l}{0.1880} & \multicolumn{1}{l}{0.3080} & \multicolumn{1}{l}{0.1410} & 0.1847 & \multicolumn{1}{l}{0.0366} & \multicolumn{1}{l}{0.1065} & \multicolumn{1}{l}{0.0482} & 0.0680 & \multicolumn{1}{l}{0.0108} & \multicolumn{1}{l}{0.0383} & \multicolumn{1}{l}{0.0105} & 0.0190 \\
\multicolumn{1}{c}{} & DeCA   & \multicolumn{1}{l}{0.1870} & \multicolumn{1}{l}{0.3076} & \multicolumn{1}{l}{0.1402}    &  0.1804    & \multicolumn{1}{l}{0.0327}    & \multicolumn{1}{l}{0.0990}    & \multicolumn{1}{l}{0.0388} & 0.0590 & \multicolumn{1}{l}{0.0103}    & \multicolumn{1}{l}{0.0381}    & \multicolumn{1}{l}{0.0101}    &  0.0182   \\
\multicolumn{1}{c}{} & BOD   & \multicolumn{1}{l}{0.1890}    & \multicolumn{1}{l}{0.3082}    & \multicolumn{1}{l}{0.1414}    &   0.1828   & \multicolumn{1}{l}{0.0375}    & \multicolumn{1}{l}{0.0134}    & \multicolumn{1}{l}{0.0489} & 0.0703 &    \multicolumn{1}{l}{\uline{0.0126}} & \multicolumn{1}{l}{\uline{0.0389}} & \multicolumn{1}{l}{0.0119} & \uline{0.0215}\\
\cmidrule(l){2-14} 
\multicolumn{1}{c}{} & $\mathsf{\textbf{DCF (Ours)}}$   & \multicolumn{1}{l}{\textbf{0.1979}*} & \multicolumn{1}{l}{\textbf{0.3134}*} & \multicolumn{1}{l}{\textbf{0.1439}*} &  \textbf{0.1853}  & \multicolumn{1}{l}{\textbf{0.0513}*} & \multicolumn{1}{l}{\textbf{0.1210}*} & \multicolumn{1}{l}{\textbf{0.0642}*} &  \textbf{0.0816}*  & \multicolumn{1}{l}{\textbf{0.0132}*} & \multicolumn{1}{l}{\textbf{0.0411}*} & \multicolumn{1}{l}{\textbf{0.0132}*} & \textbf{0.0223}*  \\ 

\midrule
\midrule
\multicolumn{1}{c}{\multirow{7}{*}{NGCF}}     & Normal & \multicolumn{1}{l}{0.1235} & \multicolumn{1}{l}{0.2257} & \multicolumn{1}{l}{0.0934} & 0.1291 & \multicolumn{1}{l}{0.0335} & \multicolumn{1}{l}{\uline{0.1015}} & \multicolumn{1}{l}{\uline{0.0452}} & 0.0634 & \multicolumn{1}{l}{0.0172} & \multicolumn{1}{l}{0.0495} & \multicolumn{1}{l}{\uline{0.0174}} & 0.0273\\
\multicolumn{1}{c}{} & WBPR   & \multicolumn{1}{l}{0.1237} & \multicolumn{1}{l}{0.2252} & \multicolumn{1}{l}{0.0936} & 0.1289 & \multicolumn{1}{l}{0.0331} & \multicolumn{1}{l}{0.1013} & \multicolumn{1}{l}{0.0446} & 0.0637 & \multicolumn{1}{l}{0.0166} & \multicolumn{1}{l}{0.0481} & \multicolumn{1}{l}{0.0164} & 0.0267\\
\multicolumn{1}{c}{} & WRMF   & \multicolumn{1}{l}{0.1244} & \multicolumn{1}{l}{0.2255} & \multicolumn{1}{l}{0.0942} & 0.1304 & \multicolumn{1}{l}{0.0334} & \multicolumn{1}{l}{0.1011} & \multicolumn{1}{l}{0.0449} & \uline{0.0639} & \multicolumn{1}{l}{0.0169} & \multicolumn{1}{l}{0.0486} & \multicolumn{1}{l}{0.0167} & 0.0270\\
\multicolumn{1}{c}{} & T-CE   & \multicolumn{1}{l}{\uline{0.1260}} & \multicolumn{1}{l}{\uline{0.2270}} & \multicolumn{1}{l}{\uline{0.0959}} & \uline{0.1313} & \multicolumn{1}{l}{\uline{0.0335}} & \multicolumn{1}{l}{0.1014} & \multicolumn{1}{l}{0.0450} & 0.0635 & \multicolumn{1}{l}{0.0173} & \multicolumn{1}{l}{\uline{0.0497}} & \multicolumn{1}{l}{0.0173} & 0.0270\\
\multicolumn{1}{c}{} & DeCA   & \multicolumn{1}{l}{0.1172}    & \multicolumn{1}{l}{0.2235}    & \multicolumn{1}{l}{0.0846}    &  0.1037  & \multicolumn{1}{l}{0.0318} & \multicolumn{1}{l}{0.0973} & \multicolumn{1}{l}{0.0436} & 0.0627 & \multicolumn{1}{l}{0.0169}    &  \multicolumn{1}{l}{0.0464}    & \multicolumn{1}{l}{0.0166}    &   0.0268  \\
\multicolumn{1}{c}{} & BOD   & \multicolumn{1}{l}{0.1212}    & \multicolumn{1}{l}{0.2246}    & \multicolumn{1}{l}{0.0901}    &   0.1265   & \multicolumn{1}{l}{0.0321}    & \multicolumn{1}{l}{0.1008}    & \multicolumn{1}{l}{0.0437} & 0.0633 & \multicolumn{1}{l}{\uline{0.0174}}    & \multicolumn{1}{l}{0.0492}    & \multicolumn{1}{l}{0.0169}    &  \uline{0.0274}   \\
\cmidrule(l){2-14} 
\multicolumn{1}{c}{} & $\mathsf{\textbf{DCF (Ours)}}$    & \multicolumn{1}{l}{\textbf{0.1267}*} & \multicolumn{1}{l}{\textbf{0.2275}**} & \multicolumn{1}{l}{\textbf{0.0970}*} & \textbf{0.1321}*  & \multicolumn{1}{l}{\textbf{0.0353}*} & \multicolumn{1}{l}{\textbf{0.1037}*} & \multicolumn{1}{l}{\textbf{0.0468}*} & \textbf{0.0647}* & \multicolumn{1}{l}{\textbf{0.0180}*} & \multicolumn{1}{l}{\textbf{0.0503}**} & \multicolumn{1}{l}{\textbf{0.0179}**} &  \textbf{0.0277}** \\ 

\midrule
\midrule
\multicolumn{1}{c}{\multirow{7}{*}{LightGCN}} & Normal & \multicolumn{1}{l}{0.1236} & \multicolumn{1}{l}{0.2257} & \multicolumn{1}{l}{0.0933} & 0.1289 & \multicolumn{1}{l}{0.0347} & \multicolumn{1}{l}{0.1029} & \multicolumn{1}{l}{0.0457} &  0.0642 & \multicolumn{1}{l}{0.0185} & \multicolumn{1}{l}{0.0514} & \multicolumn{1}{l}{0.0183} &\uline{0.0291} \\
\multicolumn{1}{c}{} & WBPR   & \multicolumn{1}{l}{0.1239} & \multicolumn{1}{l}{0.2253} & \multicolumn{1}{l}{0.0914} & 0.1295 & \multicolumn{1}{l}{0.0353} & \multicolumn{1}{l}{0.1043} & \multicolumn{1}{l}{0.0460} & 0.0638 & \multicolumn{1}{l}{0.0182} & \multicolumn{1}{l}{0.0514} & \multicolumn{1}{l}{0.0178} & 0.0282\\
\multicolumn{1}{c}{} & WRMF   & \multicolumn{1}{l}{0.1242} & \multicolumn{1}{l}{0.2260} & \multicolumn{1}{l}{0.0928} & 0.1298 & \multicolumn{1}{l}{\uline{0.0357}} & \multicolumn{1}{l}{\uline{0.1046}} & \multicolumn{1}{l}{\uline{0.0464}} & \uline{0.0650} & \multicolumn{1}{l}{0.0181} & \multicolumn{1}{l}{0.0515} & \multicolumn{1}{l}{0.0180} & 0.0283\\
\multicolumn{1}{c}{} & T-CE  & \multicolumn{1}{l}{\textbf{0.1261}} & \multicolumn{1}{l}{\uline{0.2261}} & \multicolumn{1}{l}{\uline{0.0956}} &  \uline{0.1306} & \multicolumn{1}{l}{0.0290} & \multicolumn{1}{l}{0.1020} & \multicolumn{1}{l}{0.0409} & 0.0612 & \multicolumn{1}{l}{\uline{0.0185}} & \multicolumn{1}{l}{\uline{0.0516}} & \multicolumn{1}{l}{\uline{0.0183}} & 0.0290\\
\multicolumn{1}{c}{} & DeCA   & \multicolumn{1}{l}{0.1185} & \multicolumn{1}{l}{0.2251} & \multicolumn{1}{l}{0.0859} & 0.1038 & \multicolumn{1}{l}{0.0347} & \multicolumn{1}{l}{0.0987} & \multicolumn{1}{l}{0.0440} &  0.0640 & \multicolumn{1}{l}{0.0176} & \multicolumn{1}{l}{0.0503} & \multicolumn{1}{l}{0.0172} & 0.0273\\ 
\multicolumn{1}{c}{} & BOD   & \multicolumn{1}{l}{0.1225} & \multicolumn{1}{l}{0.2254} & \multicolumn{1}{l}{0.0902} & 0.1287 & \multicolumn{1}{l}{0.0325} & \multicolumn{1}{l}{0.1014} & \multicolumn{1}{l}{0.0443} & 0.0638  & \multicolumn{1}{l}{0.0182} & \multicolumn{1}{l}{0.0513} & \multicolumn{1}{l}{0.0177} & 0.0282\\ 
\cmidrule(l){2-14} 
\multicolumn{1}{c}{} & $\mathsf{\textbf{DCF (Ours)}}$    & \multicolumn{1}{l}{\uline{0.1258}} & \multicolumn{1}{l}{\textbf{0.2274}*} & \multicolumn{1}{l}{\textbf{0.0961}**} &  \textbf{0.1315}*  & \multicolumn{1}{l}{\textbf{0.0365}*} & \multicolumn{1}{l}{\textbf{0.1050}} & \multicolumn{1}{l}{\textbf{0.0472}*} & \textbf{0.0659}*   & \multicolumn{1}{l}{\textbf{0.0192}*} & \multicolumn{1}{l}{\textbf{0.0523}*}    & \multicolumn{1}{l}{\textbf{0.0187}**} & \textbf{0.0296}**\\ 
\bottomrule[1.2pt]
\end{tabular}}
\label{performance comparison}
\end{table*}

\subsection{Experimental Settings}
\subsubsection{Datasets. }We conduct extensive comparative experiments on three public popular datasets: Adressa\footnote{https://github.com/WenjieWWJ/DenoisingRec\label{web}}~\cite{SGDL, T-CE, DeCA}, MovieLens\footnote{https://github.com/wangyu-ustc/DeCA}~\cite{SGDL, DeCA}, Yelp\textsuperscript{\ref{web}}~\cite{SGDL, T-CE}. Detailed statistics of the datasets are in Table~\ref{tab:dataset statistics}. For getting clean test datasets, we only include interactions with a dwell time of at least 10 seconds, based on~\cite{DeCA, T-CE} in Adressa. For MovieLens, we create a test set that only includes interactions with a rating of five, following the setting from ~\cite{DeCA}. Similarly, for Yelp, we refer to~\cite{T-CE} and only include interactions with a rating higher than three in our clean test set.

Note that to maintain consistency with the settings of previous denoising works~\cite{T-CE, DeCA, BOD} and ensure fair performance comparisons. All ratings and dwell times filter the clean test set, while the training and validation sets are not filtered.

\subsubsection{Evaluation protocols.}
We follow~\cite{T-CE, BOD} to split datasets into the training set, validation set, and clean test set with the ratio 8:1:1.
Following existing works on denoising recommendations~\cite{SGDL, T-CE, DeCA}, we report the results \textit{w.r.t.} two widely used metrics: NDCG@\textit{K} and Recall@\textit{K}, where higher scores indicate better performance. For a comprehensive comparison of different models, we set \textit{K}=5 and \textit{K}=20 for all datasets. Each experiment is repeated five times, as well as we conduct a significance test on the results of the experiments. In addition, limited by the page length, we place the detailed parameter setting at~\ref{Parameter}.

\subsubsection{Baselines.} Our goal in this paper is to weaken the adverse impact of noisy interactions on model performance. For this purpose, we choose four \textbf{backbones} based on implicit feedback:
\begin{itemize}[leftmargin=*]
  \item \textbf{GMF}~\cite{GMF+NeuMF}: This generalized version of matrix factorization captures the latent factors of users and items.
  \item \textbf{NeuMF}~\cite{GMF+NeuMF}: NeuMF combines matrix factorization and neural networks to enhance the accuracy of collaborative filtering. \looseness=-2
  \item \textbf{NGCF}~\cite{ngcf}: NGCF is a graph neural network model for enhanced and personalized recommendation systems.
  \item \textbf{LightGCN}~\cite{lightgcn}: LightGCN is a simplified GCN designed specifically for recommendation, leveraging the power of graph-based methods to enhance interaction predictions. 
\end{itemize}
The following approaches are used to train each model:
\begin{itemize}[leftmargin=*]
  \item \textbf{Normal}~\cite{GMF+NeuMF}: The model is being trained using the simple binary-cross-entropy (BCE) loss function.
  \item \textbf{WBPR}~\cite{WBPR}: This method considers popular but non-interactive items as true negatives based on interaction count.
  \item \textbf{WRMF}~\cite{WRMF}: WRMF design weighted matrix factorization where the weights remain fixed to remove noise from recommendation.
  \item \textbf{T-CE}~\cite{T-CE}: This is a generalized version of BCE to truncate large-loss examples with a dynamic threshold in each iteration. 
  \item \textbf{DeCA}~\cite{DeCA}: DeCA utilizes two different models to predict and account for any disagreement that may arise from noisy samples.
  \item \textbf{BOD}~\cite{BOD}: BOD automatically learns samples weights using bi-level optimization. 
\end{itemize}

\subsection{Performance Comparison (RQ1)}
To validate the effectiveness and generalization of our framework, we conduct extensive experiments on four popular backbones and three datasets.
Table~\ref{overrall compa} shows the results of our DCF with existing denoising methods. Indices that perform the \textbf{best} are in bold, while \uline{runner-ups} are underlined. According to Table~\ref{overrall compa}, we can draw the following observations and conclusions:
\begin{itemize}[leftmargin=12pt]
  \item DCF achieves good performance on all backbones and datasets. We attribute these improvements to the extended observation of sample loss values, searching for valuable hard samples, and relabeling some highly discriminated noisy samples. After relabeling, these samples can be utilized to avoid performance degradation due to sparse sample space. However, baseline models such as BOD and DeCA lack these capabilities, making them less effective than ours.
  \item We observe that T-CE is second only to us on most datasets and backbone, whereas the other DeCA and BOD are not as good as T-CE. T-CE directly drops the high loss samples, whereas DeCA and BOD do not utilize the high and low loss phenomena. The results are consistent with previous studies~\cite{T-CE, MBA}. Furthermore, DeCA is based on the prediction agreement of the two models, and BOD is based on bi-level optimization, and we speculate that the suboptimal performance because they are both unstable in training. Hence, we argue that using loss values is a straightforward and efficient method.
  \item In other baselines, the BOD is second only to us in Yelp. We think this is because the bi-level optimization of BOD requires more density data in order to better learn the weighting matrix involved. Whereas WRMF sometimes achieves good results, we speculate it is because matrix factorization has better effects on sparse data. We also note that on the MovieLens dataset, other denoising methods do not outperform normal training. We argue this is because on sparse datasets, these reweighting or dropping strategies may be less effective due to inadequate representation learning, which leads to true-positive samples being mistakenly down-weighted or dropped.
\end{itemize}

\subsection{Model Investigation (RQ2)}
\subsubsection{Ablation Study.} DCF consists of three components, Confirmed Loss Calculation~(CL), Hard Sample Search~(HS), and Progressive Label Correction~(LC). We are eager to validate the effectiveness of each component and the combination between them on the performance. Therefore, we conduct the following seven sets of ablation experiments~(as shown in Tabel~\ref{ablation study}). However, due to the length of the paper, we only show results on the MoiveLens dataset and GMF, and other datasets are similar and omitted. In addition, we also conduct ablation experiments on damping function in~\ref{Effect of Damping Function}.

We observe better results from more components, which aligns with our expectations. We analyze this trend in effectiveness as being attributable to each component playing its desired role. In the single component section, we observe the best results for $\text{DCF}_\text{HS}$. This proves that we obtain hard samples by deriving a confidence lower bound on the loss. In addition, $\text{DCF}_\text{LC}$ also achieves good performance, which is attributed to the fact that we let the corrected samples be retrained to avoid sample wasting, as well as misalignment of the training and testing sample spaces. Also, in the combination of components, HS and LC achieved excellent results, proving the effectiveness of the two-by-two combination. It is worth noting that the combination of CL and HS does not achieve more significant results. We hypothesize that the reason for this is that the search for HS is a game of risk and benefit, and therefore, we have to tune the hyperparameters carefully.

\begin{table}[t]
\caption{The effect of each components on \(\mathsf{DCF}\).}\vspace{-1em}
\tabcolsep=0.03cm
\begin{tabular}{@{}lcccc@{}}
\toprule
\textbf{Method} & \textbf{R@5} & \textbf{R@10} & \textbf{N@5} & \textbf{N@10} \\ 
\midrule
\midrule
\(\mathsf{T\text{-}CE}\)      & 0.0374           & 0.0734  & 0.0509  &  0.0591 \\
\midrule
\(\mathsf{DCF}_{\text{CL}}\)    & 0.0432(15.5\%)            & 0.0751(2.3\%)      &  0.0527(3.5\%)  &   0.0600(1.5\%)  \\
\(\mathsf{DCF}_{\text{HS}}\)     & 0.0435(16.3\%)            & 0.0772(5.2\%)  & 0.0535(5.1\%)  &  0.0610(3.2\%) \\
\(\mathsf{DCF}_{\text{LC}}\)                   &  0.0430(15.0\%)            & 0.0756(3.0\%)            & 0.0529(3.9\%) & 0.0604(2.2\%)   \\
\midrule
\(\mathsf{DCF}_{\text{CL+HS}}\)             & 0.0436(16.6\%)             & 0.0743(1.2\%)        & 0.0531(4.3\%)  & 0.0596(0.8\%)    \\
\(\mathsf{DCF}_{\text{CL+LC}}\)                   & 0.0459(22.7\%)             & 0.0763(4.0\%)           &  0.0549(7.9\%)  & 0.0612(3.6\%)  \\
\(\mathsf{DCF}_{\text{HS+LC}}\)     & 0.0454(21.4\%)            & 0.0779(6.1\%)     & 0.0545(7.1\%)  & 0.0613(3.7\%) \\
\midrule
\(\mathsf{DCF}_{\text{ALL}}\)    & \textbf{0.0471(25.9\%)} & \textbf{0.0789(7.5\%)} & \textbf{0.0553(8.6\%)} & \textbf{0.0621(5.1\%)} \\
\bottomrule
\end{tabular}
\label{ablation study}
\end{table}

\subsubsection{Parameters Sensitivity Analysis.}
\begin{figure}[t]
    \centering
    {
        \includegraphics[width=0.34\linewidth]{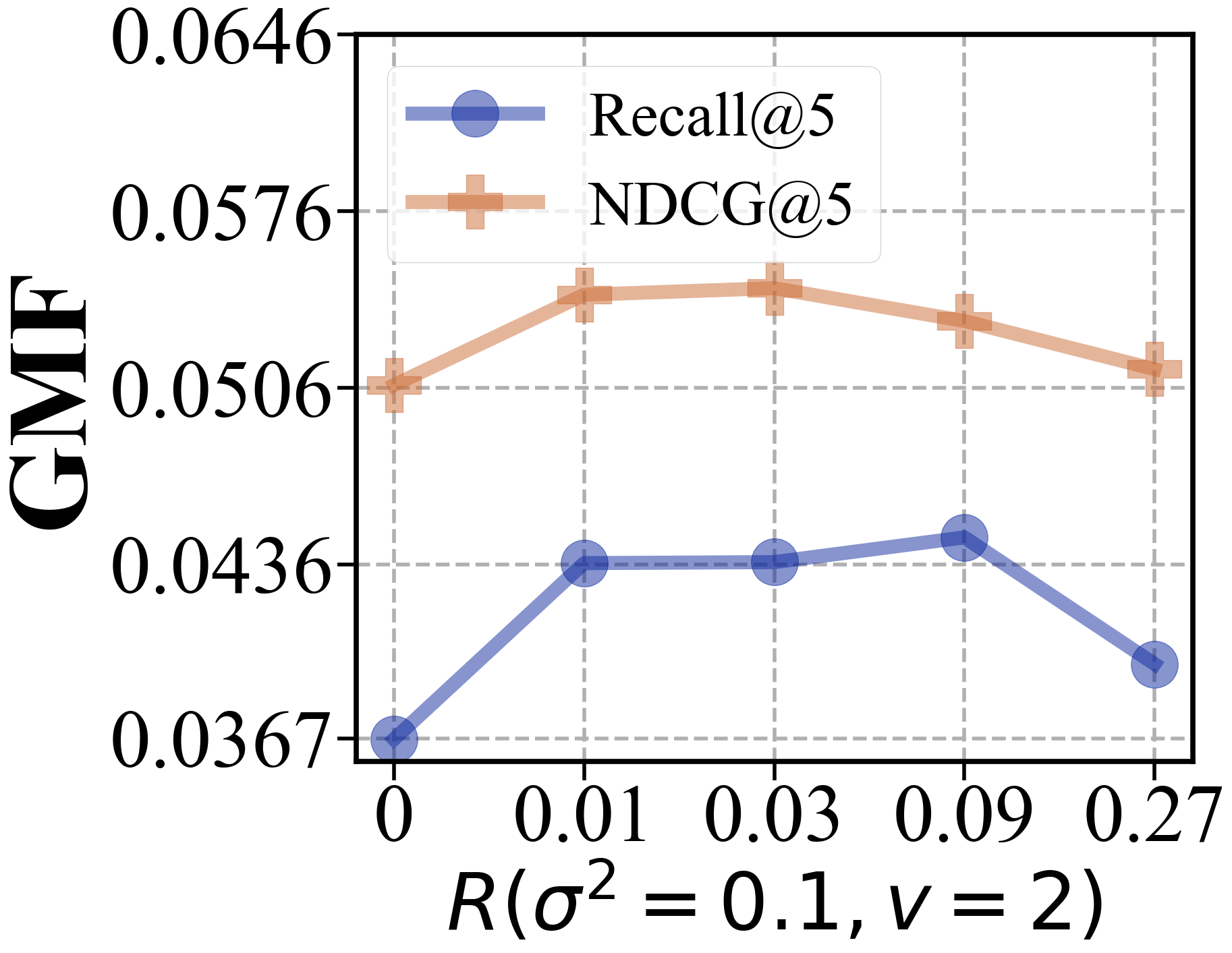}
    }
    {
        \includegraphics[width=0.305\linewidth]{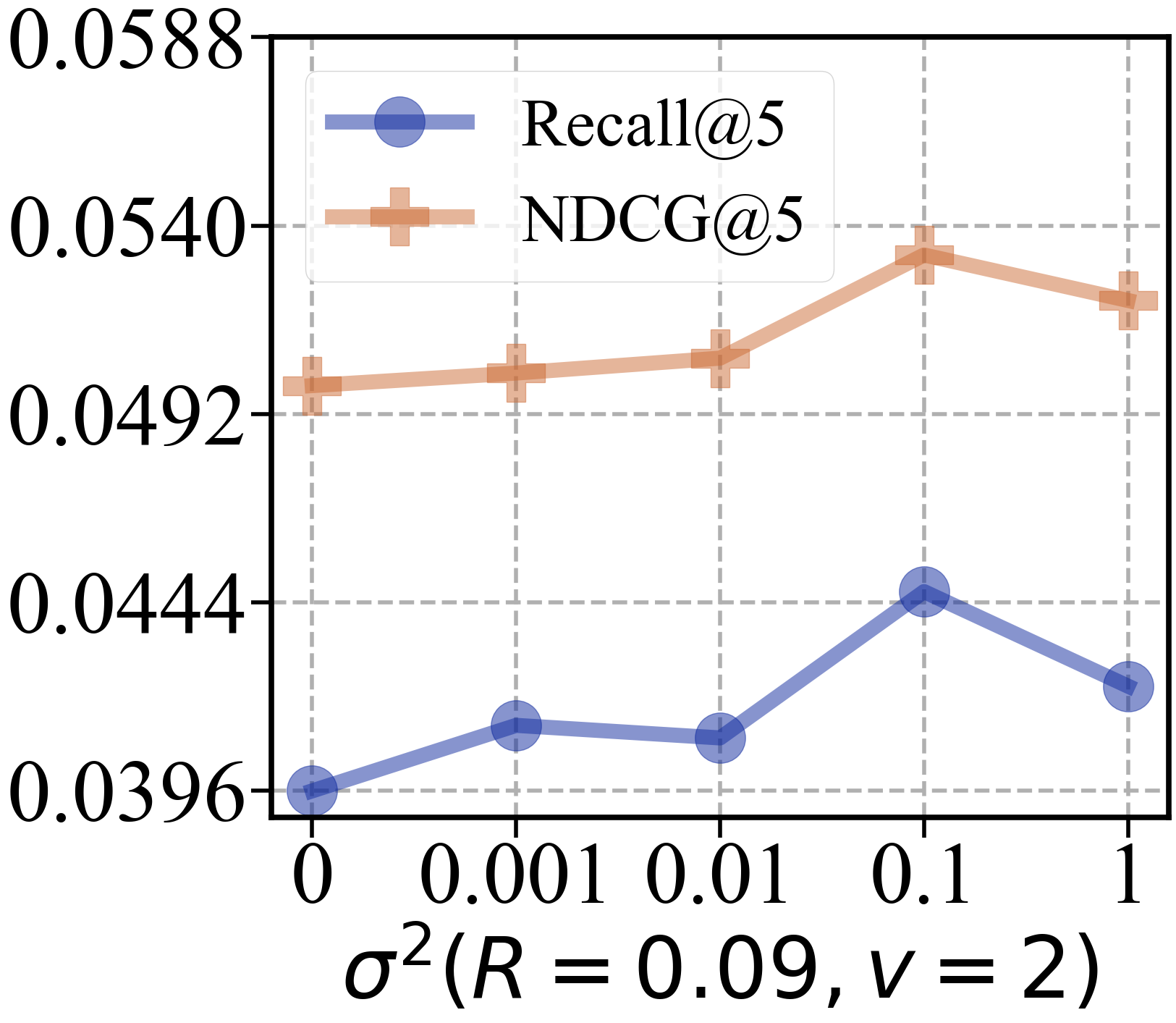}
    }
    {
        \includegraphics[width=0.305\linewidth]{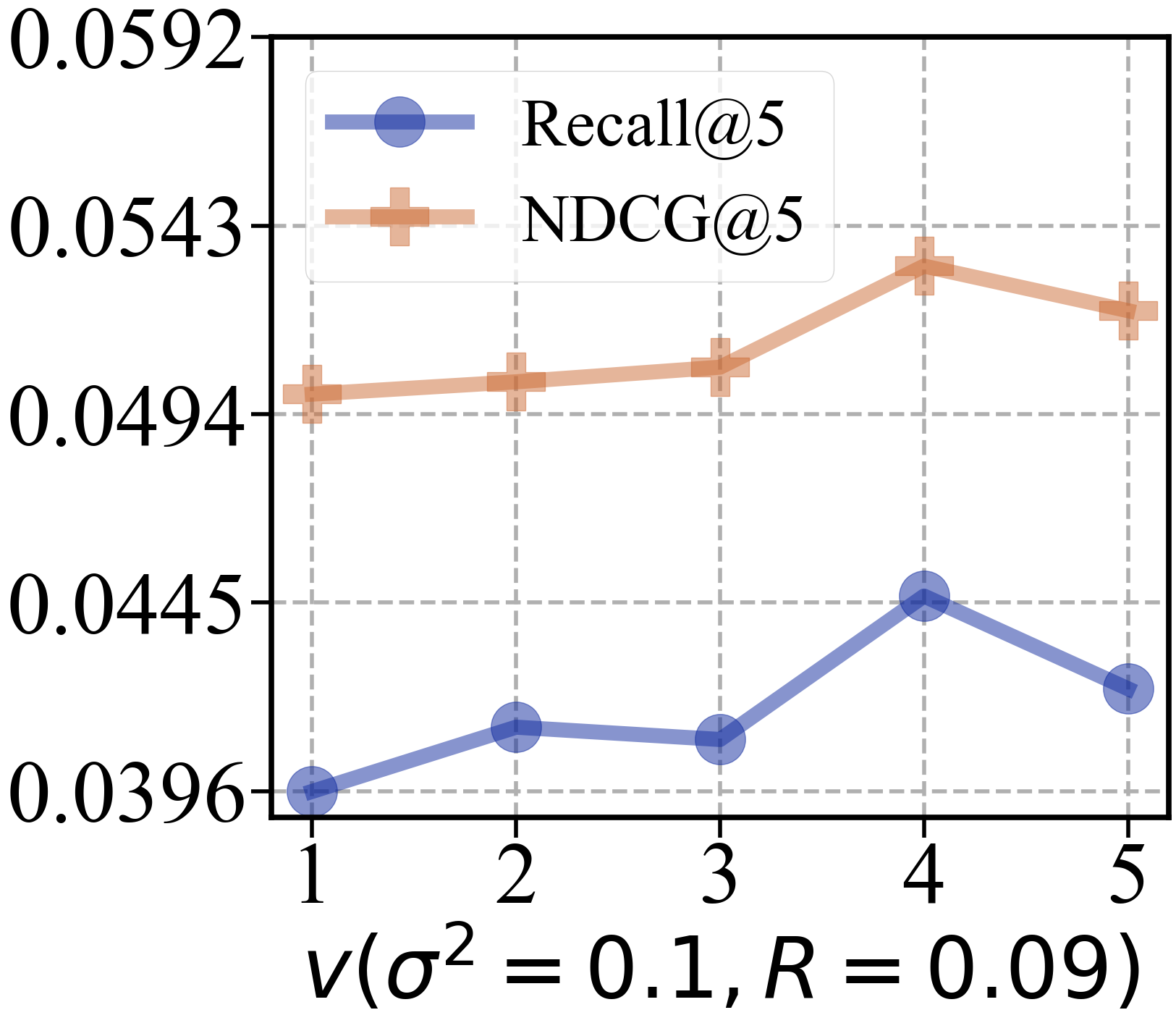}
    }\\
    {
        \includegraphics[width=0.337\linewidth]{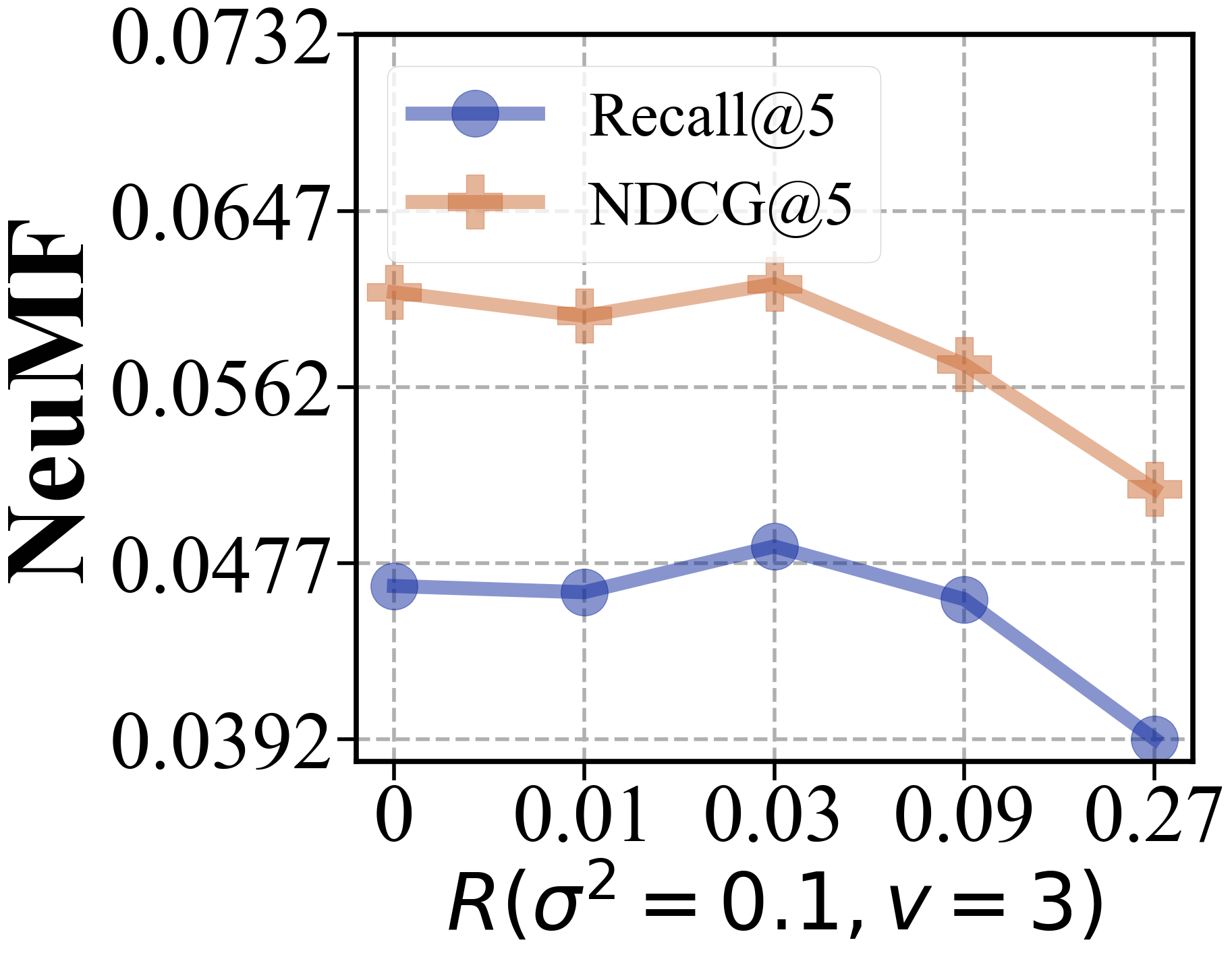}
    }
    {
        \includegraphics[width=0.305\linewidth]{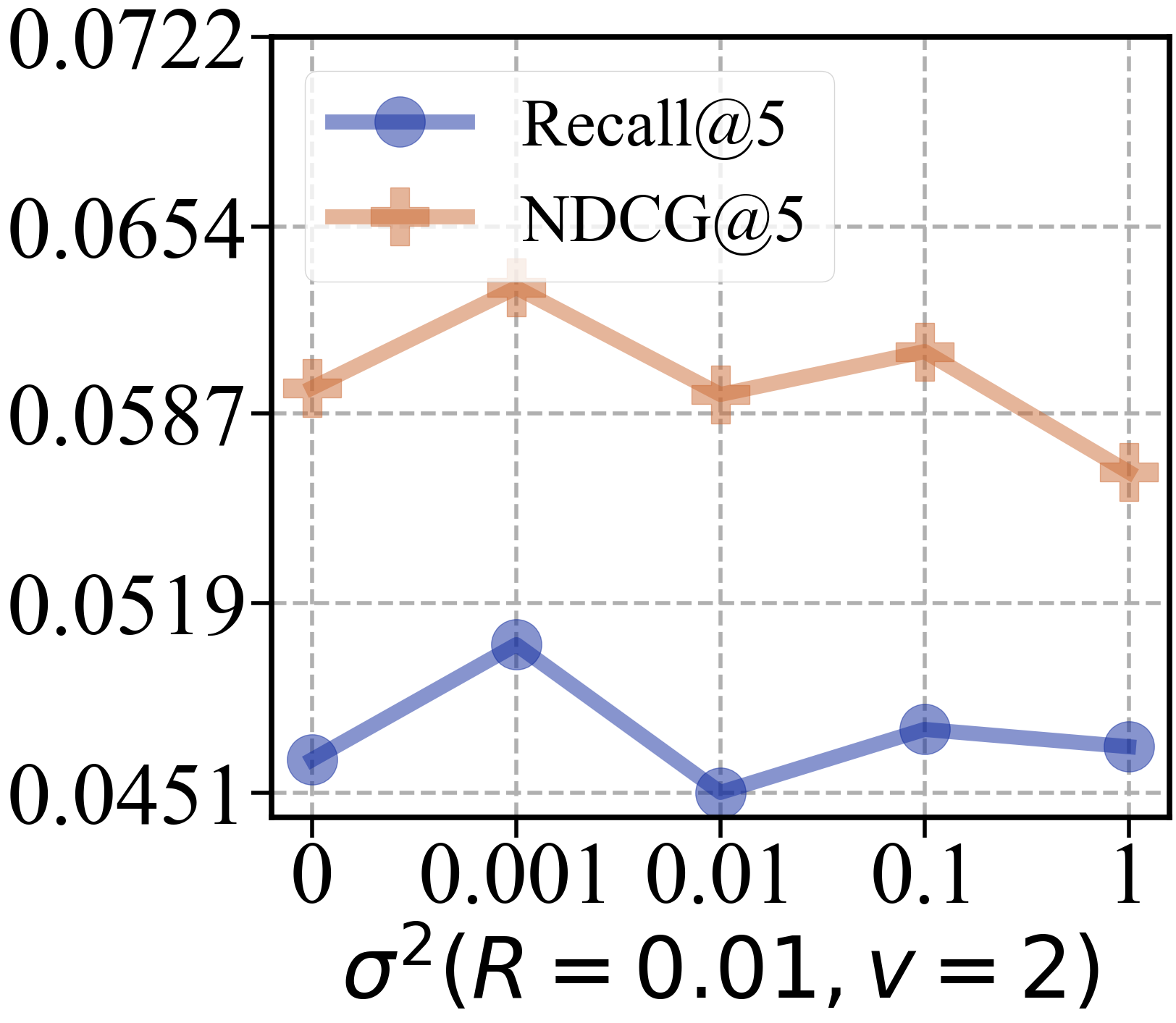}
    }
    {
        \includegraphics[width=0.305\linewidth]{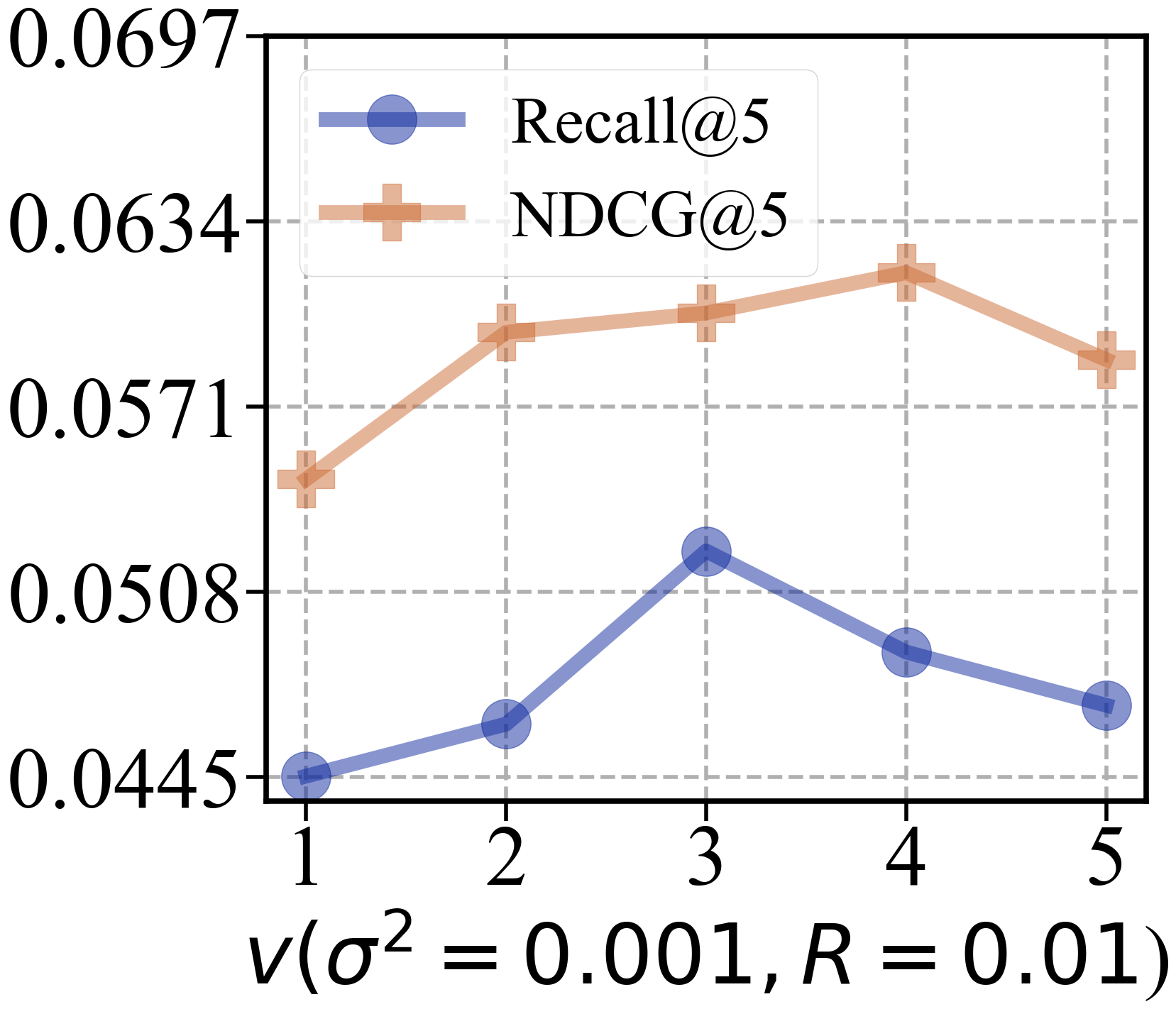}
    }
    \caption{Impact of the relabel ratio $R$, search discretion level $\sigma^2$ and time interval $v$.}
    \label{hyper}
    \vspace{-7pt}
\end{figure}

We are eager to know the sensitivity of different critical parameters: ($\text{i}$) relabeling ratio $R$; ($\text{ii}$) discretion level $\sigma^2$ of searching hard samples; ($\text{iii}$) calculating time interval $v$ of mean losses. Experiments are conducted on the Movielens dataset. As shown in Fig.~\ref{hyper}, we can find that:
\begin{itemize}[leftmargin=12pt]
  \item NDCG@5 and Recall@5 are affected by the hyperparameters, and the overall trend is up and then down. We observe that the performance is not getting better as the flip rate $R$ increases. We speculate that this is because only a tiny fraction of the highest loss samples are strongly correlated with the noisy samples, and thus more flips may lead to more errors.
  \item $\sigma^2$ controls the intensity of the search for hard samples, which is a risk versus reward game; the larger the $\sigma^2$, the higher the probability of performance degradation, but at the same time it is possible to improve the effectiveness of the model.
  \item  We observe that extending the time interval $v$ for mean loss computation has apparent benefits, but this is limited to a specific range. If the range is too wide, the calculation of the mean may be negatively affected by the instability of early model training.
\end{itemize}

\begin{figure}[t]
	\centering
    {\includegraphics[width=.49\linewidth]{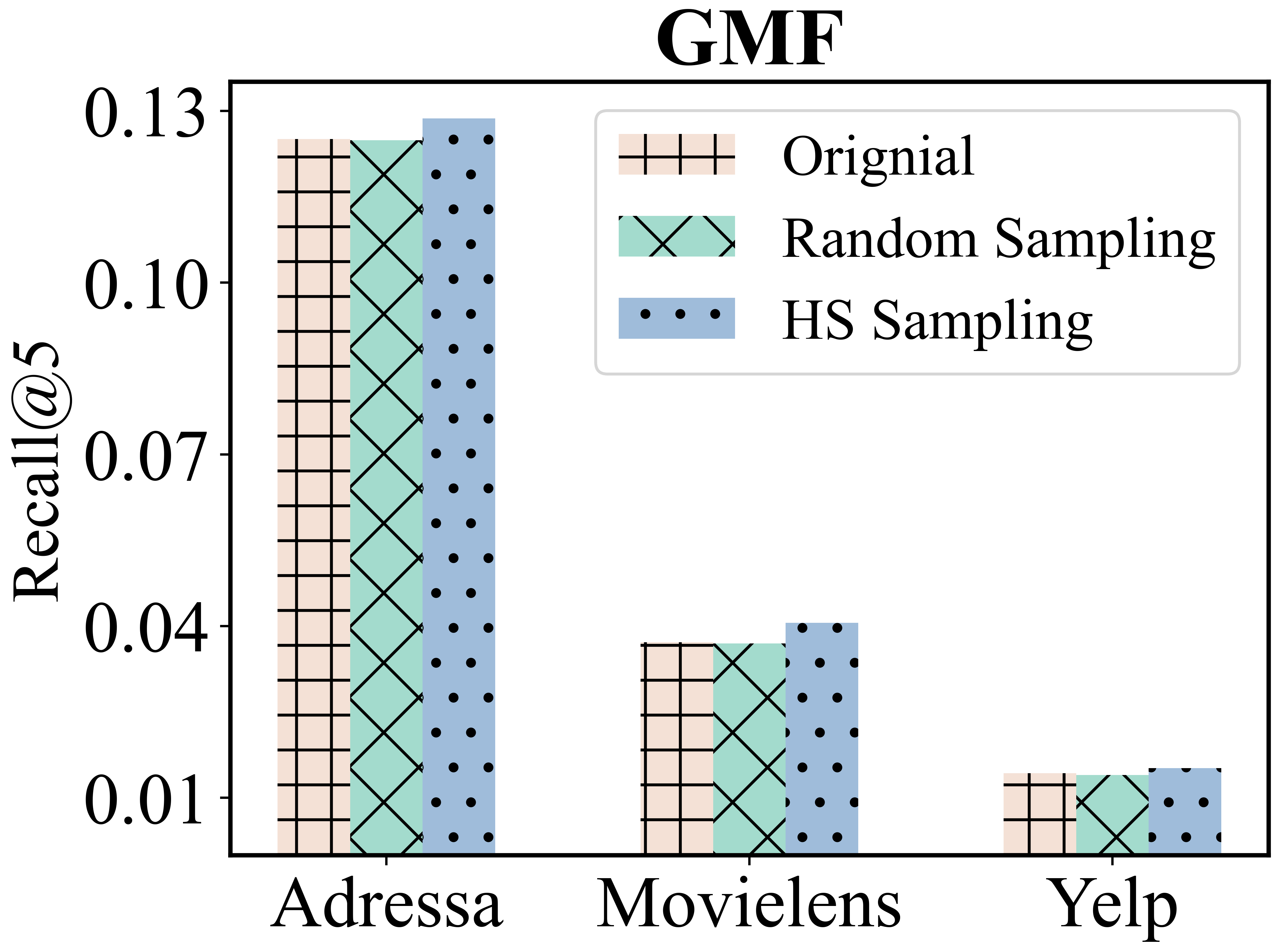}}\hspace{2pt}
    {\includegraphics[width=.49\linewidth]{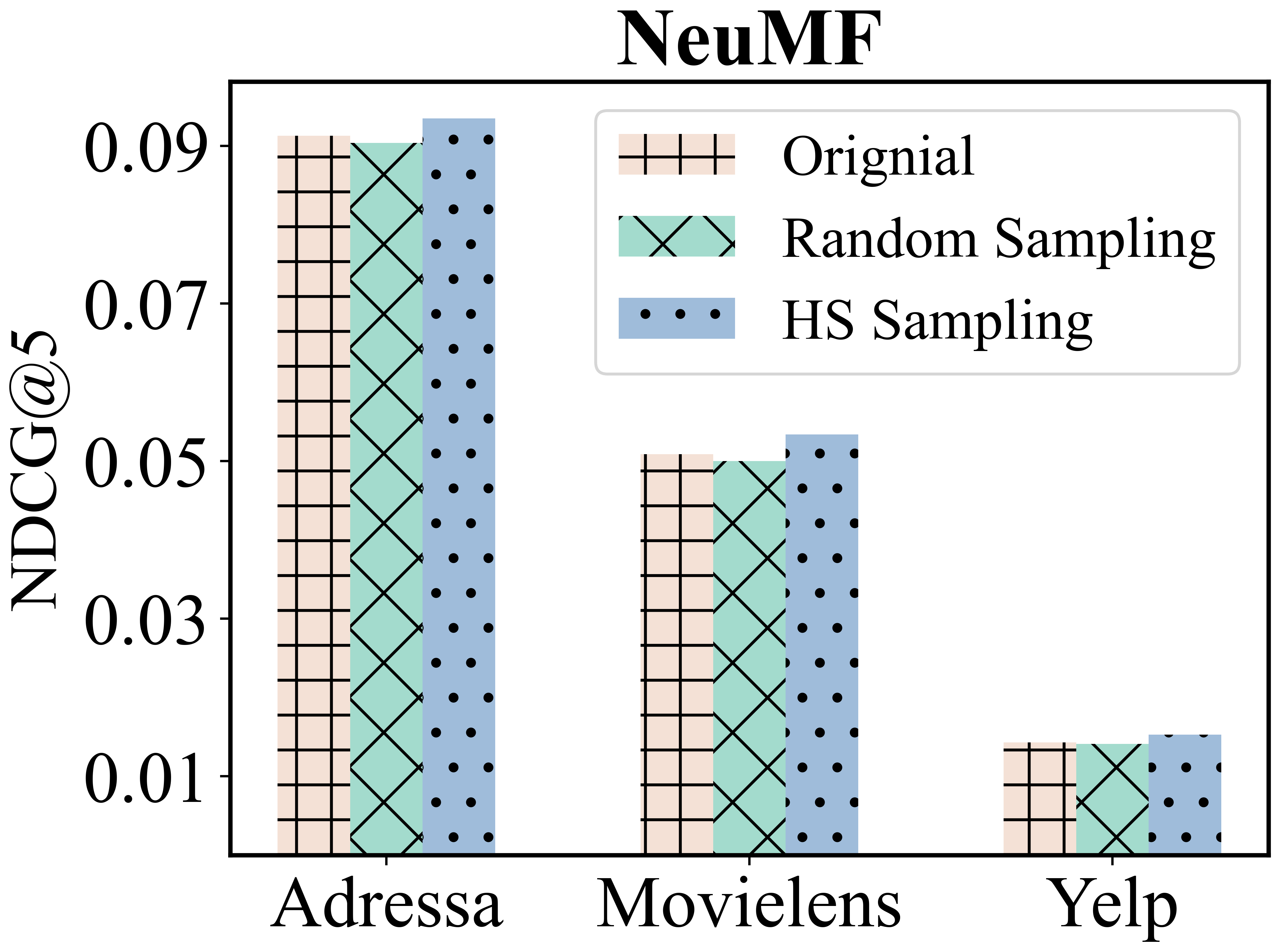}}
   \caption{Comparative experiments on three datasets with two backbones validate the effectiveness of our hard sample search strategy to improve model performance.}
\vspace{-2mm}
\label{hard}
\end{figure}

\subsection{Hard Sample Search Strategy~(RQ3)}
Despite confirming the effectiveness of hard sample search in our ablation experiments, we aim to ascertain whether the identified samples are indeed hard samples rather than noisy samples. Therefore, we design a comparative experiment. The first group is to use the searched samples for T-CE training. The second group is to randomly sample the same size of samples from the dropped samples for training, and the third group is not to use the original training of both samples to compare the performance of these three. Experiments are conducted on three datasets and two backbones.

From Fig.~\ref{hard}, we observe that adding hard samples improves model performance, which is consistent with our expectations. This is due to the efficient identification of noisy samples through our lower bound on loss. In contrast, adding random samples from the dropped samples to the model does not enhance the recommendation model performance but has a negative impact.

\subsection{Progressive Strategy in Label Correction (RQ4)}

\begin{figure}[t]
	\centering
    {\includegraphics[width=.485\linewidth]{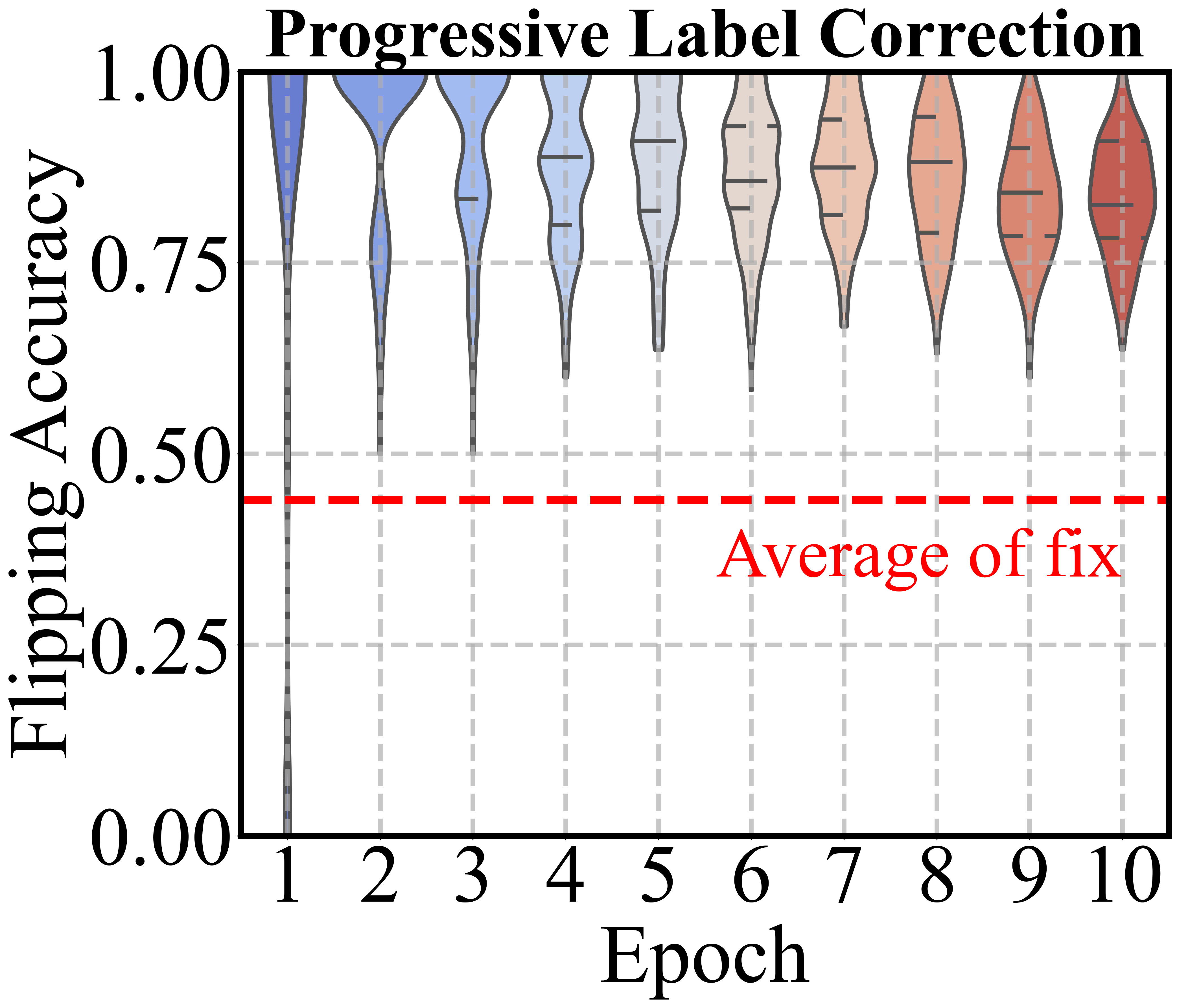}}\hspace{5pt}
    {\includegraphics[width=.485\linewidth]{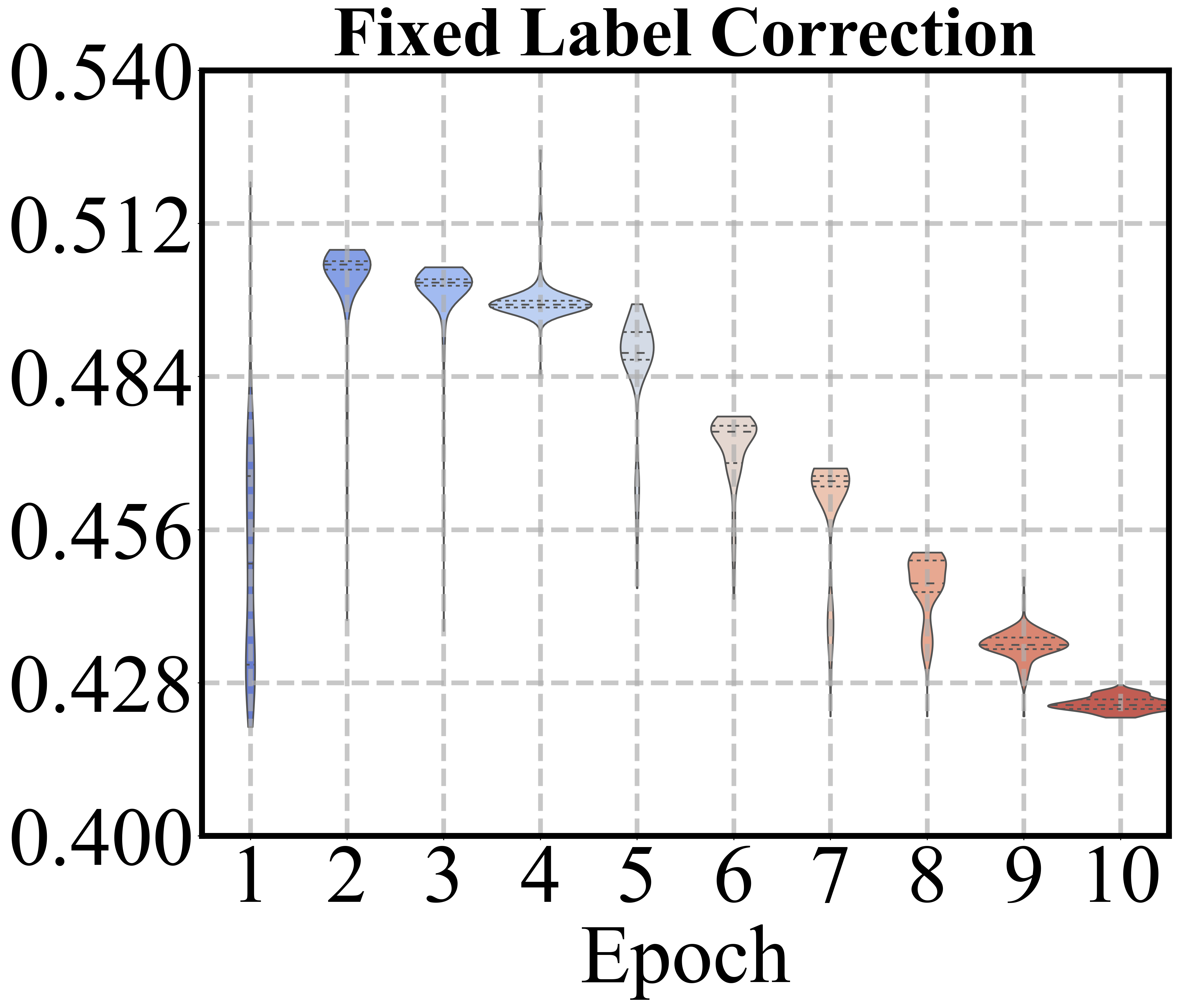}}
   \caption{Comparison of flip accuracy between progressive label correction and fixed. For a clear presentation, we use a violin plot here. Additionally, we mark the average flip accuracy of fixed with a red line to clearly highlight the superiority of our progressive strategy.}
\vspace{-2mm}
\label{flip}
\end{figure}

We have demonstrated the effectiveness of progressive label correction in the ablation section. However, we are interested in gaining a more detailed understanding of our strategy's performance, particularly regarding flip accuracy. We also wish to determine if our progressive relabeling approach is superior to fixed relabeling methods. The results of the experiment are shown in Fig.~\ref{flip}.
\begin{itemize}[leftmargin=12pt]
  \item From the two figures, it can be observed that the flip accuracy of progressive label correction is significantly higher than that of the fixed strategy~(red lines). We attribute this to our dynamic relabeling ratio strategy, better suited for transitioning from instability to stability training characteristics. In the early stages of training, the model may be more susceptible to the interference of noisy labels, which could impact its final performance. However, in the later stages, the accuracy of the progressive relabeling decreases slightly. We presume that this is because the model already has enough information, and there is a slight overfitting phenomenon.
  \item We also observe that the stability of the flip accuracy in the progressive strategy increases as training iterations progress and ultimately maintain a high level. Additionally, in the beginning, even if we flip a small proportion, the volatility is still high. If we were to flip a larger proportion, there is a high likelihood of introducing incorrect information. This further validates the effectiveness of our strategy of gradually increasing the flip ratio from low to high.
\end{itemize}

\section{Related Work}
Recommendation systems based on implicit feedback have attracted a lot of attention. However, recent studies point out implicit feedback is easily vulnerable to users' unconscious behaviors and various biases (e.g., popularity bias, position bias, etc.), which degrade the generalization ability. Thus to weaken the problem caused by noisy implicit feedback, some denoising methods~\cite{robust-survey, one_class, WRMF, Sampler_Design, knowledge_refine} have been proposed, and they can be categorized into sample drop~\cite{T-CE, AutoDenoise-reinforce, Sampler_Design, ss2, PU_learning, missing_data, user_exposure, multi-view}, sample 
reweight~\cite{T-CE, AutoDenoise-weight, BOD, WBPR}, and other methods~\cite{DeCA} designed with the help of other information. 

\vspace{5pt}
\noindent
\textbf{Drop-based Methods.} An intuitive idea is to pick out the clean interactions and send them for training. Then, distinguishing between the characteristics of clean and noisy samples is critical. T-CE~\cite{T-CE} observe experimentally loss values of noisy samples are higher than clean samples, and experiments using this observation find it effective. Subsequent studies~\cite{AutoDenoise-reinforce, SGDL} incorporate this observation into the design of denoising recommendation models. For example, AutoDenoise~\cite{AutoDenoise-reinforce} considers the denoising work to consist of two behaviors: searching and deciding. So they use reinforcement learning to automate these two behaviors. Additionally,~\cite{missing_data} proposes to augment recommendation algorithms with noisy examples of user preferences and mitigate the challenge of data sparsity.

\vspace{5pt}
\noindent
\textbf{Reweight-based Methods.} In addition to selecting clean samples, some work~\cite{BOD, T-CE, AutoDenoise-weight} reduce the weights on the noisy samples, which reduces the impact on model parameter updates and avoids the reduction of generalization ability. For example, R-CE~\cite{T-CE} utilizes the loss value as a denoising signal, thus assigning a small weight to the noisy sample. AutoDenoise~\cite{AutoDenoise-weight} also utilizes the loss value as a denoising signal and fuses it with the user representation and item representation in the weight calculation with good results. BOD~\cite{BOD} converts the learning of weights into a bi-level optimization problem and uses a simple and elegant solution to learn the weight parameters with relatively good results.

\vspace{5pt}
\noindent
\textbf{Other Methods.} First, methods inspired by the phenomenon of noisy samples during training. For example, DeCA~\cite{DeCA} observes that different models have greater predictive divergence for noisy samples and lesser divergence for clean samples. SGDL~\cite{SGDL} observe the memorization effect of noise, thus removing noise in the pre-training period. Thus based on this observation, an efficient denoising recommendation model is designed and performs well. Second, there is also some work to denoise based on graph collaborative filtering, e.g., RGCF~\cite{graph1}, RocSE~\cite{graph2}. Their method's main design idea lies in removing noisy interaction edges. RocSE~\cite{graph2} removes the noisy interactions from both graph structure denoising and contrastive learning and performs well on graph collaborative filtering. Third, some methods to denoise in specific recommendation scenarios, design denoising modules for scenarios such as movies~\cite{vedio}, music~\cite{music}, and social~\cite{socail1, socail2}. As well as exploring the removal of noisy samples in sequential recommendation~\cite{sequential1, sequential2, sequential3}.
 
\vspace{5pt}
\noindent
\textbf{Comparison with Our Method.} We argue that the loss-based method is a simple and effective approach but suffers from predictive instability due to random initialization of parameters. 
Furthermore, both the dropped sample and reweighted sample strategies based on loss values design have inherent flaws. For example, it leads to misalignment of the training and testing spaces or the need for complex methods to learn the weights.
Therefore, we consider a more robust loss value calculation as well as relabeling.

\section{Conclusion}
In this paper, we present a novel framework named DCF, designed to mitigate the adverse effects of noisy samples on the representation learning of users and items. The DCF contains two modules. The first module, sample dropping correction, achieves more stable loss estimations by calculating the mean loss value of samples and focuses on identifying and retraining hard samples. The second module, progressive label correction, we relabel samples with a higher likelihood of being noisy and reintegrate them into the learning process. Extensive experiments on widely used benchmarks and datasets demonstrate the effectiveness and generalization of our proposed framework.

\begin{acks}
This work is supported by grants from the National Key Research and Development Program of China $($Grant No.2021ZD0111802$)$ and National Science Foundation of China $($Grant No.U23B2031, No.721881011, No.U21B2026$)$.
\end{acks}

\balance
\bibliographystyle{ACM-Reference-Format}
\bibliography{DCF}

\clearpage
\appendix
\section{APPENDIX}
\subsection{Derivation OF THEOREM 1}
\label{PROOF}
According to~\cite{Catoni_2010}, we explore the upper bound $\tilde{\mu}_z^{-}$ and the lower bound $\tilde{\mu}_z^{+}$ for the mean $\tilde{\mu}_z=\frac{1}{n} \sum_{i=1}^n \phi\left(z_i\right)$ computation, as follows
\begin{equation}
    \tilde{\mu}_z^{-} \leq \tilde{\mu}_z \leq \tilde{\mu}_z^{+}.
\end{equation}
To derive a bound for $\tilde{\mu}_z$. We build an estimator of $\tilde{\mu}_z$, the formal definition as follow,
\begin{equation}
r\left(\tilde{\mu}_z\right)=\sum_{i=1}^n \phi\left[\alpha\left(z_i-\tilde{\mu}_z\right)\right]=0 .
\end{equation}
where $\alpha$ is positive real parameter. We then adjust $r(\tilde{\mu}_z)$ to be in the form of quantity
\begin{equation}
r(\theta)=\frac{1}{\alpha n} \sum_{i=1}^n \phi\left[\alpha\left(z_i-\theta\right)\right], \theta \in \mathbb{R}.
\end{equation}
With the exponential moment inequality~\cite{Giné_Latała_Zinn_2000} and the $\mathrm{C}_r$ inequality~\cite{Mohri_Rostamizadeh_Talwalkar_2012}, we analysis the $r(\tilde{\mu}_z)$ in the form of inequalities 
\begin{equation}
    \begin{aligned}
    \mathbb{E}\{\exp [\alpha n r(\theta)]\} & \leq \left\{1+\alpha\left(\mu_z-\theta\right)+\frac{\alpha^2}{2}\left[\alpha^2+(\mu_z-\theta)^2\right]\right\}^n \\
    & \leq \exp \left\{n \alpha\left(\mu_z-\theta\right)+\frac{n \alpha^2}{2}\left[\alpha^2+(\mu_z-\theta)^2\right]\right\}.
    \end{aligned}
\end{equation}
In order to bound $\mu_z$, we will find two non-random values $\theta_{-}$ and $\theta_{+}$ of the parameter such that with large probability $r\left(\theta_{-}\right)>0>r\left(\theta_{+}\right)$, which will imply that $\theta_{-}<\widehat{\theta}_\alpha<\theta_{+}$, since $r \left(\widehat{\theta}_\alpha\right)=0$ by construction and $\theta \mapsto r(\theta)$ is non-increasing.
The new bounds are as follow
\begin{equation}
B_{-}(\theta)=\mu_z-\theta-\alpha\left[\sigma^2+\left(\mu_z-\theta\right)^2\right]-\frac{\log \left(\epsilon^{-1}\right)}{\alpha n}
\end{equation}
and
\begin{equation}
B_{+}(\theta)=\mu_z-\theta+\alpha\left[\sigma^2+\left(\mu_z-\theta\right)^2\right]+\frac{\log \left(\epsilon^{-1}\right)}{\alpha n} .
\end{equation}
From~\cite{21stat} (Lemma 2.2), we obtain that
\begin{equation}
P\left(r(\theta)>B_{-}(\theta)\right) \geq 1-\epsilon \quad \text { and } \quad P\left(r(\theta)<B_{+}(\theta)\right) \geq 1-\epsilon .
\end{equation}
According to Chebyshev's inequality and the previous proposition~\cite{21stat}. We assume
\begin{equation}
4 \alpha^2 \sigma^2+\frac{4 \log \left(\epsilon^{-1}\right)}{n} \leq 1.
\end{equation}
From~\cite{21stat} (Theorem 2.6), we then have
\begin{equation}
\tilde{\mu}_z^{-} \geq \mu_z-\frac{\alpha \sigma^2+\frac{\log \left(\epsilon^{-1}\right)}{\alpha n}}{\alpha-1}
\end{equation}
and
\begin{equation}
\tilde{\mu}_z^{+} \leq \mu_z+\frac{\alpha \sigma^2+\frac{\log \left(\epsilon^{-1}\right)}{\alpha n}}{\alpha-1} .
\end{equation}
With probability at least $1-2 \epsilon$, we have $\tilde{\mu}_z^{-} \leq \tilde{\mu}_z \leq \tilde{\mu}_z^{+}$. We set $\alpha=\frac{n}{\sigma^2}$, and obtain
\begin{equation}
\left|\tilde{\mu}_z-\mu_z\right| \leq \frac{\sigma^2\left(n+\frac{\sigma^2 \log \left(\epsilon^{-1}\right)}{n^2}\right)}{n-\sigma^2}.
\end{equation}
With probability of at least $1-2\epsilon$, we proceed to exploit the lower bound by setting $\epsilon=\frac{1}{2i}$.
$$
\ell^{\ast}=\tilde{\mu}-\frac{\sigma^2\left(i+\frac{\sigma^2 \log (2 i)}{i^2}\right)}{d-\sigma^2},
\label{low bound}
$$
where $d$ represents the number of times a sample has not been dropped.

\subsection{Parameter Settings}
\label{Parameter}
All of the models use the Adam optimizer with a learning rate of 0.001, batch size set to 1024, and embedding dimension set to 32. The number of graph convolution layers for LightGCN is set to 3 without dropout. For every training example, we select one observed interaction and one random unobserved interaction to input into the model. The ratio of relabeled interactions is tuned in \{0, 0.01, 0.03, 0.09, 0.27\}. $\sigma^2$ is tuned in \{0, 0.001, 0.01, 0.1\}. The time step is tuned in \{1, 2, 3, 4, 5\}. Note that the purpose of using hyperparameters $\sigma^2$ is to lessen the dependence on strong assumptions, helping our framework work better in practice. Our experiments are based on RTX 2080 Ti GPU and PyTorch. 

\subsection{Additional Experiments}
\subsubsection{Effect of Damping Function} 
\label{Effect of Damping Function}
We verify whether damping function~(DF) is effective through comparative experiments. By observing Table~\ref{ae_df}, we conclude that DF produces a positive effect. And we reason this is because DF removes the negative impact of outliers, thus making confirmed loss calculation more accurate.
\begin{table}[h]
\caption{The effect of damping function with GMF on \(\mathsf{DCF}\).}\vspace{-1em}
\tabcolsep=0.08cm
\begin{tabular}{lllllll}
\toprule 
\textbf{Dataset} & \textbf{Method} & \textbf{R@5} & \textbf{R@10} & \textbf{N@5} & \textbf{N@10} \\
\midrule
\midrule
\multicolumn{1}{l}{\multirow{2}{*}{Adressa}} & DCF & 0.1296 & 0.2183 & 0.0938 & 0.1254\\
& DCF w/o DF & 0.1292 & 0.2178 & 0.0933 & 0.1251\\
\multicolumn{1}{l}{\multirow{2}{*}{MovieLens}} & DCF & 0.0427 & 0.1175 & 0.0543 & 0.0743\\
& DCF w/o DF & 0.0423 & 0.1172 & 0.0540 & 0.0739\\
\multicolumn{1}{l}{\multirow{2}{*}{Yelp}} & DCF & 0.0155 & 0.0458 & 0.0158 & 0.0257\\
& DCF w/o DF & 0.0151 & 0.0453 & 0.0154 & 0.0252\\
\bottomrule
\end{tabular}
\label{ae_df}
\end{table}
\end{document}